\documentclass[aps,11pt,floatfix]{revtex4-2}
\usepackage{graphicx}
\usepackage{dcolumn}
\usepackage{bm}
\usepackage{xcolor, soul}
\usepackage{amsfonts}
\usepackage{booktabs}
\usepackage{siunitx}
\usepackage{longtable}
\usepackage{array}
\usepackage{setspace}
\usepackage{float}
\usepackage{nameref} 
\usepackage[colorlinks,citecolor=red,urlcolor=blue,bookmarks=false,hypertexnames=true]{hyperref}
\newcommand{\AS}[1]{{\color{black}{#1}}}

\newcommand{\HM}[1]{{\color{blue}{#1}}}
\newcolumntype{P}[1]{>{\centering\arraybackslash}p{#1}}
\newcolumntype{M}[1]{>{\centering\arraybackslash}m{#1}}

\begin{document}
\singlespacing

\title{\AS{A Study of Non-linear Flows and Shear Banding in Wormlike Micelles Under Varying Elasticity, Flow Curvature, and Surfactant Chemistry}}

\author{Alfredo Scigliani}
\author{Hadi Mohammadigoushki}
\email[Corresponding Author: ]{hadi.moham@eng.famu.fsu.edu}
\affiliation{Department of Chemical and Biomedical Engineering, FAMU-FSU College of Engineering, Tallahassee, FL 32310, USA.}
\affiliation{National High Magnetic Field Laboratory, Florida State University, Tallahassee, FL, USA 32310}
\date{\today}
\vspace{0.5cm}

\begin{abstract}
\vspace{0.3cm}
\setlength{\parskip}{12pt}
We report the flow dynamics of two shear-banding wormlike micellar solutions with distinct surfactant chemistries in a Taylor–Couette (TC) setup following a startup shear. The solutions, formulated with CTAB/NaSal and CPyCl/NaSal, exhibit comparable bulk rheology and equilibrium microstructural properties. By varying the TC gap size, we systematically examine elasticity number over a range of $1.28\times10^5$–$4.47\times10^6$, while flow curvature spans from $0.022$ to $0.171$. Under a step shear into the stress plateau, both solutions exhibit a pronounced stress overshoot that intensifies with increasing elasticity number and flow curvature, followed by the development of growing flow heterogeneities. Beyond a critical threshold of elasticity number and curvature, the CTAB/NaSal solution exhibits transient flow reversal, whereas the CPyCl/NaSal solution, despite developing similar heterogeneities, does not undergo flow reversal under any tested conditions. We hypothesize that this difference arises from the stronger electrostatic repulsions between CPyCl head groups due to their larger size, coupled with the higher dissociation rate of Cl$^-$ counter-ions. Additionally, the quasi-steady velocity profiles are significantly influenced by elasticity number and flow curvature. Wall slip at the outer cylinder exhibits a non-monotonic dependence on these parameters: it is negligible at low values of elasticity number and flow curvature, peaks at intermediate values, and diminishes at high values. Our findings highlight how surfactant chemistry, elasticity number, and flow curvature control shear-banding flows in wormlike micelles.

\end{abstract}

\maketitle

\section{Introduction}
{Surfactants are amphiphilic molecules possessing both hydrophilic (water-attracting) and hydrophobic (water-repellent) components. This dual nature enables surfactants to spontaneously self-assemble in aqueous solutions, forming micelles beyond a critical micelle concentration~\cite{cates2006rheology, dreiss2007wormlike, chu2013smart}. In the presence of a salt (or a counter-ion) or at higher surfactant concentrations, these molecules may self-assemble to a variety of nano-structures (e.g., rodlike, wormlike micelles and/or vesicles) depending on their packing parameter, chemistry, temperature, pH, surfactant and salt concentration, and other parameters~\cite{israelachvili2011intermolecular}. Wormlike micelles (WLMs) are elongated, flexible micellar structures that can entangle, resulting in viscoelastic behavior similar to polymer solutions. However, unlike polymers, WLMs may break and reform, providing them with additional stress relaxation mechanisms, making them useful for various applications, such as in the oil and gas industry, household products, drag reduction, and bio-medicine~\cite{dreiss2007wormlike, yang2002viscoelastic, rothstein2020complex}.\par 

Under certain conditions, these systems exhibit a flow phenomenon known as shear banding, where the fluid undergoing shear deformation separates into regions with distinct shear rates. This behavior is typically observed as a stress plateau in steady shear flow curves over a range of imposed shear rates ~\cite{olmsted2008perspectives, divoux2016shear}. At low imposed shear rates, and below the stress plateau, WLMs behave like Newtonian fluids. However, beyond a critical shear rate, the fluid can separate into coexisting high and low shear rate bands under a common shear stress~\cite{rehage1991viscoelastic, cates2008shearing}. WLM solutions are easy to prepare and are not susceptible to mechanical degradation, making them a model system for shear banding studies ~\cite{berret2005, rassolov2022kinetics, lerouge2010shear, berret1997inhomogeneous, rehage1991viscoelastic}. Most experimental studies on shear banding in WLMs have utilized flow geometries such as cone-and-plate~\cite{boukany2008use, britton1997two,casanellas2015spatiotemporal, ravindranath2008banding} and Taylor-Couette flows (TC; flow between two concentric cylinders) ~\cite{britton1999transition,mair1996observation,mair1997shear,brown2011changing}. \AS{TC is considered a canonical geometry due to its relatively simple flow profile and reduced susceptibility to complications such as edge fracture or interfacial instabilities, though these can still occur under specific conditions\cite{hu2010steady}}. Previous studies in shear banding WLMs have focused on two main aspects: long-term quasi-steady flows~\cite{fardin2012instabilities, fardin2012shear,mohammadigoushki2017inertio,mohammadigoushki2016flow,britton1999transition,mair1996observation, helgeson2009rheology,salmon2003velocity,lettinga2009competition,becu2004spatiotemporal, masselon2010influence} and/or transient evolution of flows toward shear banding~\cite{miller2007transient,hu2005kinetics, mohammadigoushki2019transient, rassolov2022kinetics}. \AS{In this context, we use the term quasi-steady to describe a state in which large-scale, macroscopic quantities such as the shear stress at the wall have reached an apparent steady value, while finer-scale features of the flow, such as the velocity profile or local band structure, may still be evolving slowly or exhibiting minor fluctuations.} Research on quasi-steady flows has explored competition between shear banding and wall slip~\cite{lettinga2009competition,salmon2003velocity}, the impact of flow instabilities on quasi-steady shear banding~\cite{fardin2012instabilities, fardin2012shear, mohammadigoushki2017inertio, mohammadigoushki2016flow}, and the microstructural evolution under shear banding conditions~\cite{helgeson2009rheology, gurnon2014spatially,feindel2010anomalous}. \par

Additionally, several other studies have examined the transient evolution of the flow field toward shear banding, particularly following the inception of flow~\cite{miller2007transient,lettinga2009competition,becu2004spatiotemporal, hu2005kinetics, al2018rheo}. Hu and Lips investigated the temporal evolution of a flow of a shear banding wormlike micelle based on cetylpyridinium chloride/sodium salicylate (CPyCl/NaSal) in a TC geometry following the imposition of flow at shear rates within the stress plateau, and showed that the shear stress response initially overshoots and then settles into a steady stress, meanwhile the velocity also relaxes to a steady shear banded profile wherein kinks are observed~\cite{hu2005kinetics}. In another study, Miller and Rothstein employed a TC geometry with high-resolution optical techniques to explore shear-banding in wormlike micellar solutions of the CPyCl/NaSal system. They found that beyond a critical stress, the system exhibits a shear stress plateau and forms distinct shear bands, a propagating damped elastic wave, fluctuations in shear-band evolution on short timescales, and a diffuse birefringent region before shear-banding onset was observed~\cite{miller2007transient}. Similar flow‑induced birefringence in flows of wormlike micelles was reported by other researchers~\cite{kim2000effects, helgeson2009rheology, osaki1979flow, manneville2008recent}. More recently, Mohammadigoushki and co-workers studied the transient evolution of flow profiles and shear stress in shear banding WLMs based on cetyltrimethylammonium bromide/sodium salicylate (CTAB/NaSal) following a step shear flow in a TC geometry. The experiments revealed a stress overshoot followed by a decay to a steady state and a pronounced elastic recoil, characterized by velocities opposite to the direction of imposed flow. This elastic recoil or transient flow reversal was attributed to the significantly larger elasticity number of the micellar solution~\cite{mohammadigoushki2019transient}. Later studies focused on assessing the link between elasticity number, flow ramp-up rate, and micellar entanglement density with kinetics of shear banding flow formation~\cite{rassolov2020effects, rassolov2022kinetics,rassolov2023role}. Calabrese and co-workers also investigated the evolution of heterogeneous flow and flow reversal in highly elastic gel-like WLMs formed with triblock poloxamer solution~\cite{mccauley2024heterogeneity,mccauley2023evolution}. \\

The transient evolution of flows toward shear banding in WLMs has been extensively investigated using theoretical models, which are generally classified into two fundamental groups. Mesoscopic models, such as the Vasquez–Cook–McKinley (VCM)~\cite{vasquez2007network, zhou2014wormlike, pipe2010wormlike}, Germann–Cook–Beris (GCB)~\cite{germann2014investigation, germann2013nonequilibrium}, and Bautista–Manero~\cite{bautista2002irreversible}, incorporate flow induced micellar breakage and reformation in their governing equations. The VCM model, in particular, has been successful in replicating various aspects of shear banding observed in WLMs~\cite{zhou2014wormlike, pipe2010wormlike, zhou2012multiple}. It conceptualizes micelles as Hookean dumbbells that can break into shorter lengths and recombine, predicting phenomena such as transient flow reversal above a critical elasticity number $El$~\cite{zhou2014wormlike}. Here, the elasticity number is defined as the ratio of viscoelastic forces to those of inertia: 
\begin{equation}
 El = \frac{\lambda \eta_0}{\rho d^2}.
\end{equation}
Here \( \eta_0 \) is the zero shear viscosity of the fluid, \( \lambda \) is the relaxation time, \( \rho \) is the fluid density, and \( d \) denotes the gap size of the TC cell. The second group of models approximate the flow of WLMs solutions as polymer solutions, such as the diffusive Rolie–Poly (DRP)~\cite{adams2011transient}, diffusive Johnson–Segalman models~\cite{fielding2007complex, adams2008interplay}, diffusive Giesekus model \cite{helgeson2009rheology}, a two-species Rolie-Poly and reactive-rod model\cite{salipante2024two}. According to the DRP model, kinetics of shear banding flow formation is critically dependent on micellar entanglement density \( Z \) and TC curvature $q = \ln\left(\frac{R_o}{R_i}\right)$, where \( R_o \) and \( R_i \) are the outer and inner cylinder radii, respectively. Recently, Varchanis et al.~\cite{varchanis2022evaluation} provided a comprehensive analysis by comparing several widely-used constitutive models, including Johnson–Segalman (JS), Giesekus (GS), and the VCM against experimental data from both simple shear and complex, mixed shear-extensional flow geometries. \AS{Quantitative comparisons have identified some limitations in the VCM model’s ability to predict complex fluid flows. While the model often captures the qualitative trends observed in experiments, it does not always achieve quantitative agreement, particularly in extensional flow regimes. In certain cases, the VCM model has been found to overestimate extension rates at high flow strengths and to produce stress contours that differ from those measured experimentally~\cite{varchanis2022evaluation}.}\\

The relationship between the kinetics of shear banding flow formation in WLMs and elasticity number, as predicted by the VCM model and recently observed by Rassolov et al.~\cite{rassolov2020effects} warrants further investigation. Previous experimental studies have adjusted the elasticity number by varying the rheological properties of the wormlike micellar fluid through variation of surfactant and salt concentrations~\cite{rassolov2020effects}. In particular, the salt and surfactant concentrations were changed by more than one order of magnitude. Altering surfactant/salt concentrations may significantly influence the spatiotemporal evolution of the shear banding flows in WLMs. For instance, as surfactant concentration increases, the solution may transition from semidilute to concentrated, and even approach the isotropic-to-nematic (I-N) transition, for which the flow can trigger a shear-induced I-N transition, impacting the microstructure and consequently altering the flow profile evolution. Although a comprehensive phase diagram for CTAB/NaSal wormlike micellar solutions is not available, it is plausible that surfactant concentration variations in that study could affect flow profile development. Alternatively, the effect of the elasticity number on transient flow behavior can be investigated by independently varying the gap size of the TC cell, without altering the surfactant or salt concentrations. This approach ensures that the observed effects are directly attributable to the elasticity number, isolating it from other influencing factors reported in previous studies. Furthermore, numerical simulations by Adams et al. \cite{adams2011transient}, based on the DRP fluid model, predict that the kinetics of shear-banding flow formation and the phenomenon of transient flow reversal are significantly influenced by the curvature of the TC cell. Specifically, their findings suggest that as the TC cell curvature increases, both transient shear banding and transient flow reversal become more pronounced. The latter theoretical predictions have yet to be experimentally tested in flows of WLMs.\\

{Surfactant chemistry, including the size of the polar head group, tail length, and counter-ion type, is well-known to influence the rate of micellar self-assembly and, consequently, the bulk rheological properties of micellar solutions \cite{lutz2016viscoelasticity, lutz2017intermicellar}. Lutz-Bueno et al. demonstrated that for micellar solutions comprising two surfactants with distinct chemistries (e.g., CTAB and CPyCl), the addition of sodium salicylate at constant salt and surfactant concentrations resulted in notable differences in self-assembled structures and corresponding rheological behaviors. The key differences between these surfactants lie in the size of their head groups and the type of counter-ions (Cl$^{-}$ versus Br$^{-}$). Lutz-Bueno et al. suggested that the CPyCl head group is larger than that of CTAB, leading to stronger electrostatic repulsions between the surfactant head groups. Consequently, it is expected that at a fixed sodium salicylate concentration, complete charge screening may not occur for CPyCl, which leads to weaker bulk rheological properties~\cite{lutz2016viscoelasticity, lutz2017intermicellar}. While previous studies have explored and demonstrated the formation of quasi-steady shear-banding flows in various semi-dilute surfactant systems with differing chemistries (e.g., CPyCl/NaSal \cite{britton1997two, mair1997shear, zhang2018flow, miller2007transient, gaudino2017effect, gaudino2015adding, cheng2017distinguishing, salmon2003velocity, lettinga2009competition, al2018rheo, al2020characterization, hu2005kinetics, gurnon2014spatially, britton1999transition, lopez2004shear, lopez2006rheo, feindel2010anomalous, fardin2012shear,rassolov2022kinetics}, CTAB/NaNO$_3$ \cite{brown2011changing, lerouge2008interface, mohammadigoushki2016flow}, CTAB/NaSal \cite{helgeson2009rheology, decruppe2006local, mohammadigoushki2019transient, rassolov2020effects, rassolov2023role} or other ones \cite{calabrese2015rheology,calabrese2016understanding, thareja2011shear,arenas2020alignment}), the influence of surfactant chemistry on the kinetics of shear-banding flow formation and the flow reversal phenomenon remains largely unexplored under otherwise identical bulk rheological and flow cell conditions. Prior investigations have primarily studied the kinetics of shear-banding flow formation for wormlike micelles that exhibit different rheological properties (e.g., zero-shear viscosity, width of the stress plateau, plateau modulus, relaxation time) and microstructural characteristics (e.g., micelle length, micelle entanglement density), as well as under different flow cell configurations (e.g., Taylor-Couette cell curvature or gap size). A key question that will be addressed in this study is: Do variations in surfactant chemistry influence the kinetics of shear-banding flow formation, particularly with respect to transient flow heterogeneities such as flow reversal, when bulk rheological properties, microstructural characteristics, and flow conditions are held constant? 
Recent studies have provided evidence that shear banding can be driven or strongly modified by spatial variations in micelle concentration and breakage kinetics, features that ultimately depend on the surfactant chemistry. \AS{In particular, two-fluid continuum frameworks, notably  developed for polymer solutions, that solve coupled momentum and species-balance equations have demonstrated that flow-concentration coupling alone can drive compositional heterogeneity and generate shear banding, even for fluids exhibiting a monotonic flow curve~\cite{cromer2014study, peterson2016shear}.} Most existing continuum-based models do not account for the role of surfactant chemistry in their flow analysis. Therefore, understanding the extent to which surfactant chemistry impacts the transient evolution of flows toward a shear-banding state could provide critical insights to refine and enhance theoretical frameworks for shear banding in WLM systems.\\

{The primary objective of this paper is to investigate the macroscale kinetics of shear-banding flow formation in WLMs by addressing three critical factors that may influence this complex phenomenon. First, we probe the role of the elasticity number by systematically varying the TC cell gap size ($d$), while rigorously maintaining other bulk rheological and microstructural properties as well as a fixed TC cell curvature. Second, we will experimentally test the predictions of the DRP fluid model on the effects of TC cell curvature on the kinetics of shear banding flow formation. Finally, we examine the influence of surfactant chemistry on the kinetics of shear banding by employing two distinct surfactants, CTAB and CPyCl, while carefully controlling other variables such as the elasticity number $El$, $q$, bulk rheological properties, and self-assembled microstructural characteristics.}\\

\section{Materials and Methods:}
{Two shear‐banding wormlike‐micellar solutions based on cetylpyridinium chloride (CPyCl) and cetyltrimethylammonium bromide (CTAB) were prepared with sodium salicylate (NaSal) in deionized water. Both surfactants (CTAB and CPyCl) and NaSal were obtained from Sigma-Aldrich and used as received. The rheological properties of the micellar solutions were characterized using a commercial Anton-Paar MCR 302 rheometer with an off-the-shelf concentric-cylinder geometry and custom-made TC cell geometries, which are discussed later. Small-amplitude oscillatory shear (SAOS) experiments were performed to measure the elastic and loss moduli as a function of frequency. A frequency sweep was carried out at a controlled temperature over angular frequencies ranging from 100 rad s$^{-1}$ to 0.001 rad s$^{-1}$, with eight points per decade. By analyzing the frequency response, we estimated the longest relaxation time and the micellar entanglement density. Steady-shear experiments were also conducted to obtain the quasi-steady flow curves of the wormlike-micellar solutions, with the applied shear rate ($\dot{\gamma}$) systematically varied from 0.001 s$^{-1}$ to 100 s$^{-1}$.\\

To assess the kinetics of shear‐banding flow formation, we employed custom-built rheo-optical setups consisting of a diverged 532 nm laser beam illuminating a plane orthogonal to the TC cell rotation axis and a high-speed camera (Phantom Miro 310) that recorded images of tracking particles (Potters 110P8 glass microspheres, mean diameter 8 $\mu$m) moving with the fluid. These particles, added at 50 ppm, did not alter the rheology of the WLM solutions, as shown in Fig.~S1 of the supplemental information. The recorded video was processed using TrackPy-based Python scripts to extract time-dependent local velocity data; further details are available in our prior publications~\cite{rassolov2020effects,rassolov2022kinetics,rassolov2023role}. Finally, to evaluate the impact of TC cell geometry, we fabricated a range of configurations with varying gap widths and curvature. Table~\ref{Table: RheoPTV_Cells} shows two sets of TC cells used in this study: set 1 varied the gap width $d$, and set 2 varied the curvature $q$.\par

\begin{table}[hthp]
    \centering
    \begin{tabular}{c@{\hspace{0.7cm}}c@{\hspace{0.6cm}}c@{\hspace{0.6cm}}c@{\hspace{0.6cm}}c}
        \hline\hline
        No. & $D_i$ [mm] & $D_o$ [mm] & $d$ [mm]  & $q$ \\ \hline\hline
         & 92.96 & 101.17 & 4.11 &  0.085 \\
       \textbf{set 1}  & 50.55 & 55.02 & 2.23 &  0.085 \\
         & 26.72 & 29.08 & 1.18 &  0.085 \\
         & 15.86 & 17.26 & 0.70 &  0.085 \\
        \hline        
         & 98.81 & 101.17 & 1.18 &  0.024 \\
         & 52.66 & 55.02 & 1.18 &  0.044 \\
        \textbf{set 2} & 26.72 & 29.08 & 1.18 &  0.085 \\
         & 14.90 & 17.26 & 1.18 &  0.147 \\
         & 12.64 & 15.00 & 1.18 & 0.171 \\
        \hline
    \end{tabular}
    \caption{Geometrical dimensions of different TC cells to assess the effect of the elasticity number $El$ (set 1) and the effect of TC curvature $q$ (set 2) independently. Here $D_i$, $D_o$, $d$, and $q$ refer to the inner cylinder diameter, outer cylinder diameter, the gap size, and flow curvature, respectively.}
    \label{Table: RheoPTV_Cells}
\end{table}

\section{Results and Discussion}
\subsection{Fluid Design and Characterization}
To design WLMs of different surfactant chemistry with similar rheological and self-assembled microstructure (linear worm-like chains), we created a library of wormlike micellar solutions across a range of surfactant and salt concentrations, as well as temperatures. Fig.~\ref{fig:rheology}(a,b) present the zero shear viscosity and the longest relaxation time of the micellar solutions based on CTAB and CPyCl as a function of salt to surfactant concentration ratio. Among these systems, two wormlike micelles of CTAB/NaSal (20 mM/15 mM) and CPyCl/NaSal (25 mM/20 mM) show rheological properties that are closely matched to each other (as indicated by arrows in Fig.~\ref{fig:rheology}(a,b)). Here, the longest relaxation time ($\lambda$) is obtained by fitting the multi-mode Maxwell model to the frequency sweep data. The chosen solutions are formulated such that their zero-shear viscosity remains well below the maximum zero-shear viscosity, ensuring that both primarily consist of linear micellar structures~\cite{wu2021linear,rassolov2020effects}. Fig.~\ref{fig:rheology}(c) displays the storage and loss moduli as a function of angular frequency for these two systems. In addition, Fig.~\ref{fig:rheology}(d,e) present the measured quasi-steady shear stress as a function of the imposed Weissenberg number ($Wi = \lambda \dot{\gamma}$) for these two wormlike micellar solutions, obtained across various TC cell gap sizes. The gap width of the TC cell does not affect the bulk rheological properties of these two wormlike micelles. Note that the elasticity number for these two micellar systems varies as 1.28$\times 10^5 - 4.47\times10^6$ for these TC cell gap sizes (also denoted as set 1 in Table~(\ref{Table: RheoPTV_Cells})). The bulk rheological properties of these two systems are summarized in Table~(\ref{Table:Rheology}), demonstrating that the rheological characteristics of the systems, based on two different surfactant chemistries, are similar to one another. Here the micellar entanglement density ($Z$) is estimated using the scaling relationship developed by Larson and co-workers: $Z^{0.82} \approx 3.15\frac{G'_{\text{min}}}{G''_{\text{min}}}$ where \( G'_{\text{min}} \) and \( G''_{\text{min}} \) denote the elastic and loss moduli at the frequency where \( G'' \) exhibits its local minimum~\cite{tan2021determining}. Lastly, we estimated the breakage and reptation times for both wormlike micellar systems to evaluate the dimensionless parameter $\zeta = \tau_{br}/\tau_{rep}$, which quantifies the ratio of micelle breakage time to reptation time. For CTAB, $\tau_{br} \approx 0.401$ s and $\tau_{rep} \approx 5784$ s, yielding $\zeta \approx 6.93 \times 10^{-5}$. For CPyCl, $\tau_{br} \approx 0.407$ s and $\tau_{rep} \approx 7441$ s, giving $\zeta \approx 5.47 \times 10^{-5}$. These results confirm that both systems lie firmly within the fast-breaking limit ($\zeta \ll 1$).}

\begin{figure}[hthp]
\centering
\includegraphics[width = 1\textwidth]{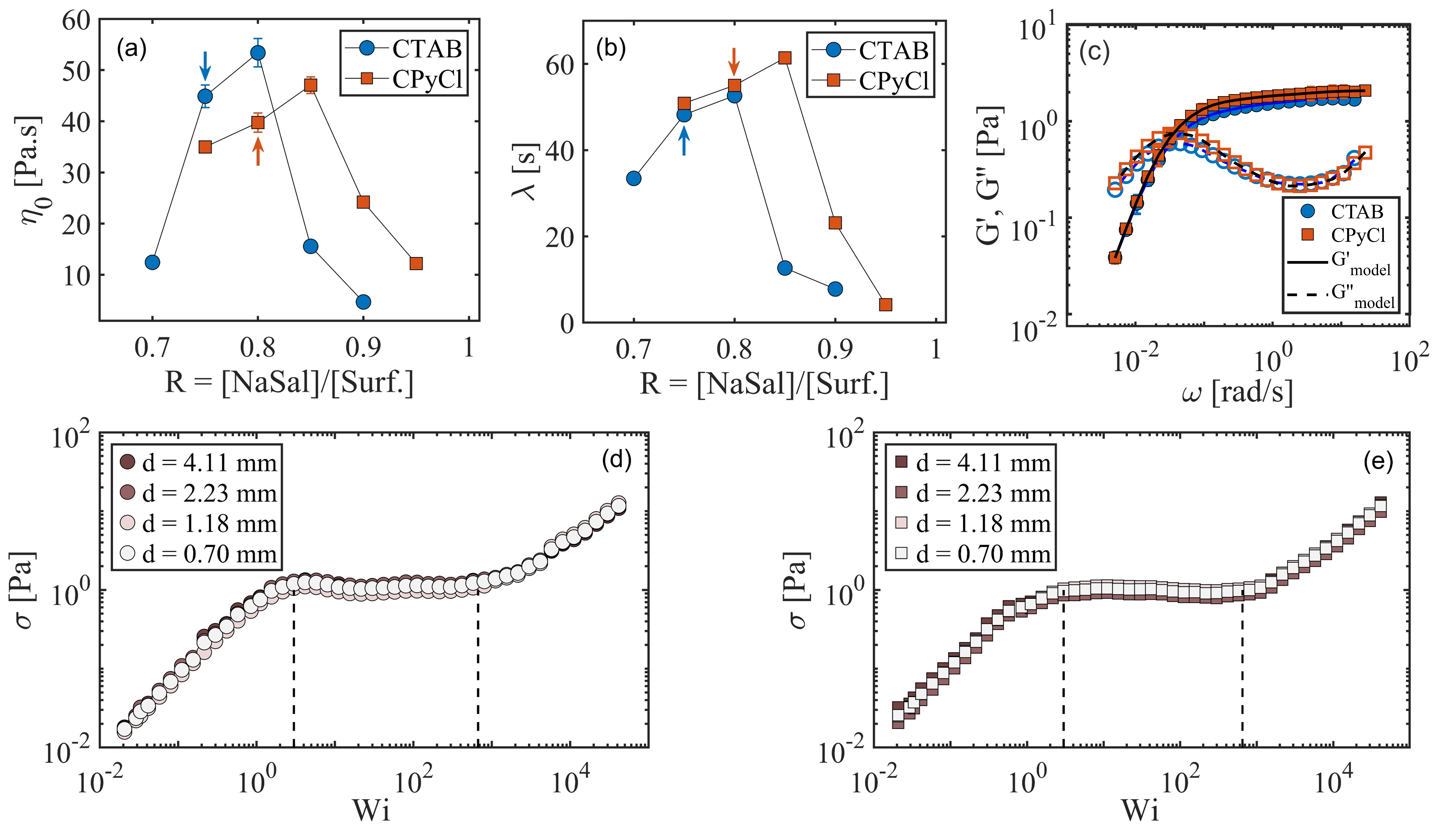}
\caption{(a) Zero-shear viscosity as a function of salt to surfactant concentration ratio. (b) Longest relaxation time as a function of salt to surfactant concentration. Arrows in (a,b) denote the selected systems for this study. (c) Storage and loss moduli as a function of angular frequency for the two selected WLM systems with matched rheological properties. (d) and (e) Quasi-steady flow curves for wormlike micellar solutions based on CTAB and CPyCl, respectively, for multiple gap sizes. Dashed vertical lines correspond to $Wi$ numbers at the onset of shear banding plateau ($Wi_{L}$) and the end of stress plateau ($Wi_{H}$).}
\label{fig:rheology}
\end{figure}

\begin{table}[H]
    \centering
    \begin{tabular}{c@{\hspace{0.6cm}}c@{\hspace{0.5cm}}c@{\hspace{0.5cm}}c@{\hspace{0.6cm}}c@{\hspace{0.6cm}}c@{\hspace{0.6cm}}c@{\hspace{0.6cm}}c@{\hspace{0.6cm}}c}
        \hline
        System & $C_\mathrm{Surf.}$ [mM] & $C_\mathrm{Salt}$ [mM] & $T$ [$^\circ$C] & $\eta_o$ [Pa$\cdot$s] & $\lambda$ [s] & Z & $Wi_{L}$ & $Wi_{H}$ \\ \hline
        CTAB/NaSal & 20 & 15 & 34.0 & 44.7 $\pm$ 2.19 & 48.2 $\pm$ 1.52 & 54 $\pm$ 3.7 & 3 & 416 \\
        CPyCl/NaSal & 25 & 20 & 20.5 & 39.9 $\pm$ 1.87 & 55.0 $\pm$ 2.02 & 59 $\pm$ 3.6 & 3 & 578 \\ \hline
    \end{tabular}
    \caption{A summary of the bulk rheological and microstructural properties of the two wormlike micellar systems selected for this study.}
    \label{Table:Rheology}
\end{table}

\subsection{Kinetics of Shear banding Flow Formation}
\subsubsection{Effect of Elasticity Number and Surfactant Chemistry} In this section, we will provide a detailed discussion on the impact of the elasticity number on various transient flow features across a range of Weissenberg ($Wi$) numbers. These features include the temporal evolution of the bulk shear stress, the degree of flow heterogeneities, the extent of transient flow reversal, wall slip (at the inner or outer cylinder), and the spatiotemporal evolution of velocity profiles. Therefore, the comparison will focus on Weissenberg numbers below the stress plateau, within the middle of the stress plateau, and approaching the end of the stress plateau. In addition, we will examine the role of surfactant chemistry by comparing the results for each surfactant system at a given $Wi$ number. \\

\begin{figure}[hthp]
\includegraphics[width=0.95\textwidth]{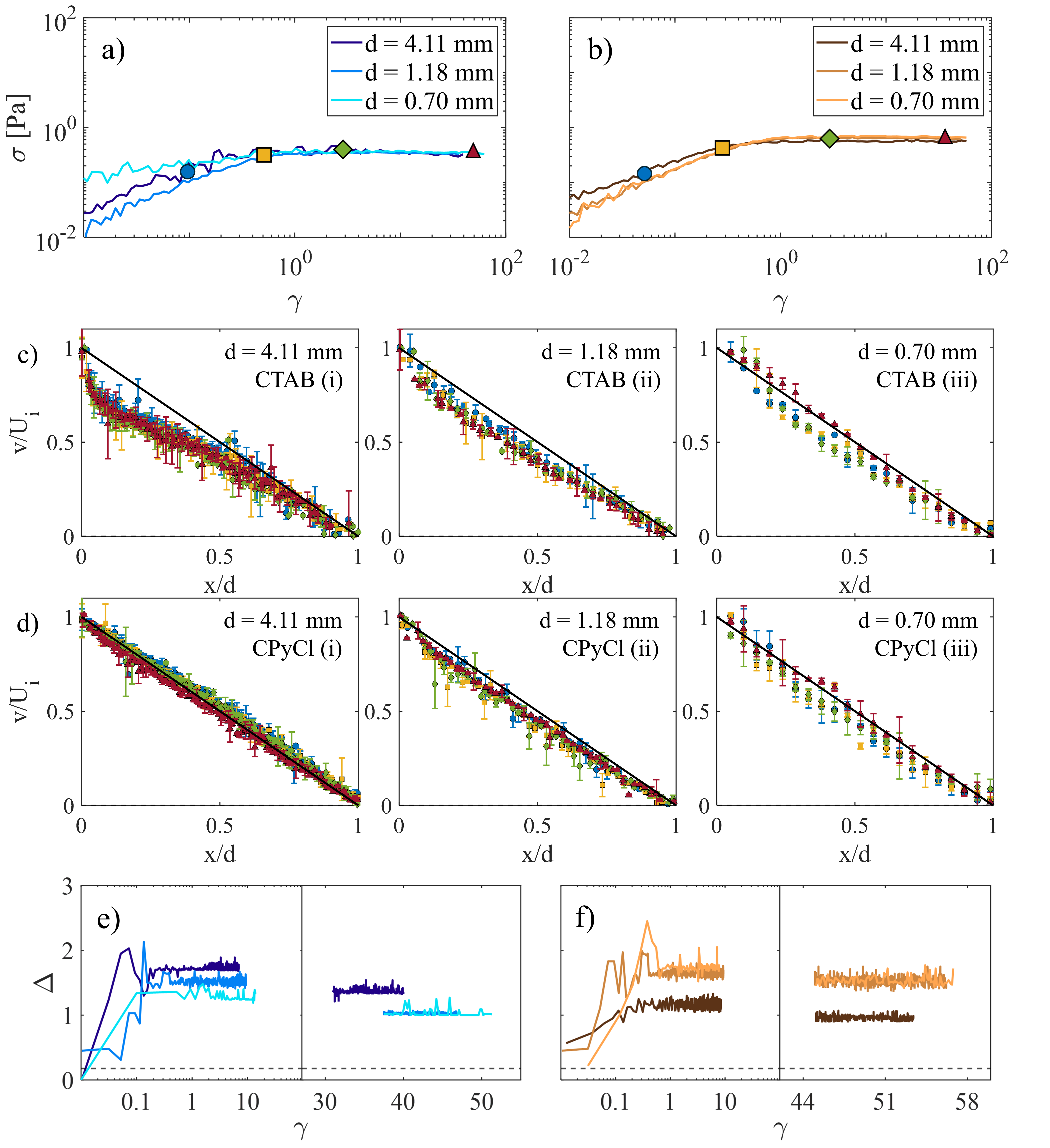}
\centering
\caption{Temporal evolution of shear stress for CTAB (a) and CPyCl (b) systems at $Wi$ = 0.5. The 1D velocity profiles are shown for various gap sizes for CTAB (c) and CPyCl (d) systems. The symbols in (a,b) shear stress curves correspond to those plotted on the corresponding (c,d) velocity profiles. The temporal evolution of flow inhomogeneities is shown for CTAB (e) and CPyCl (f) systems at the TC cell gap sizes shown in (a) and (b). The dashed line in (e,f) show the $\Delta$ value for a Newtonian fluid.}
\label{fig: E_Wi0.5}
\end{figure}

{Fig.~\ref{fig: E_Wi0.5}(a,b) display the evolution of the shear stress as a function of strain $\gamma$ at a relatively low Weissenberg number of $Wi$ = 0.5 for the two micellar systems at various gap sizes (or elasticity number). Here $\gamma$ is the strain, which is defined as $\gamma = \dot{\gamma} t$, and is used as a dimensionless representation of time. In addition, $\dot{\gamma}$ is the imposed shear rate at the inner cylinder of the TC cell. For the two micellar systems, the shear stress data across different gap sizes appear to overlap, indicating that the temporal evolution of the shear stress does not depend on the gap size (or elasticity number) below the onset of shear banding. The local velocity profiles at various points in time show slight differences across different gap sizes for the CTAB system. However, in the CPyCl-based system, the normalized 1D velocity profiles are similar to each other at various gaps. To assess the degree of flow inhomogeneity in these experiments, we assess the variations in shear rate across different gaps using $\Delta$:

\begin{equation}
\Delta = \frac{\dot{\gamma}_{\text{max}} - \dot{\gamma}_{\text{min}}}{\dot{\gamma}}.
\label{flow_inhomo}
\end{equation}

Here $\dot{\gamma}_{\text{max}}$, and $\dot{\gamma}_{\text{min}}$ are maximum and minimum shear rates within the gap of the TC cell. To compute $\Delta$, local shear rates were obtained by fitting the experimental velocity profiles to a global sigmoidal function, allowing analytical differentiation. This fit yields the high and low shear rate slopes used to compute $\Delta$, following Adams et al. \cite{adams2011transient}. As a consistency check, a piecewise-linear bin-based regression was also performed, which reproduced the same $\Delta$ within uncertainty. Full details are available in the SI of our previous work \cite{rassolov2022kinetics} and a sample computation in Fig.~S2 of the supplementary information. Fig.~\ref{fig: E_Wi0.5}(e,f) shows the temporal evolution of $\Delta$ as a function of gap size for the two micellar systems and various TC cell gap sizes. In general, $\Delta$ starts at low values and increases as the flow approaches quasi-steady and eventually levels off. It appears that the degree of flow inhomogeneities at this $Wi$ does not significantly change over the range of TC cell gaps (or fluid elasticities). In addition, $\Delta$ values at the quasi-steady regime for both CTAB and CPyCl systems are below 2, suggesting minimal inhomogeneity in the flow field. These results suggest that neither surfactant chemistry nor elasticity number affects the kinetics of flow formation for $Wi$ numbers below the stress plateau range.}\\

{Fig.~\ref{fig: E_Wi100}(a,b) illustrate the temporal evolution of the bulk shear stress for two wormlike micellar systems at $Wi$ = 100, a regime that lies within the stress plateau. For both systems and across various gap sizes, the bulk shear stress exhibits a prominent overshoot, followed by a decay toward a quasi-steady state. This stress overshoot behavior is observed similarly in both micellar solutions. Notably, the stress overshoot displays a weak dependence on elasticity number at $Wi$ = 100. As the elasticity number increases (or as the TC cell gap size decreases), the stress overshoot increases moderately for both CTAB and CPyCl micellar systems at $Wi = 100$. This finding is consistent with simulations of the VCM model, where a moderate increase in stress overshoot was observed at $Wi$ = 10 as the elasticity number was varied from 100 to infinity~\cite{zhou2014wormlike}. As the imposed $Wi$ number increases, the stress overshoot becomes more sensitive to the elasticity number in both the CTAB and CPyCl systems, as seen in Fig.~S3 of the supplementary materials. Similar to polymer melts and solutions, the overshoot in stress components in wormlike micellar solutions is linked to the micelles' resistance to chain stretching upon the onset of rapid flow. This resistance arises as the micellar structures attempt to accommodate the imposed deformation, leading to a transient buildup of stress. As the flow progresses, the micelles gradually align with the flow direction, reducing resistance to deformation and causing the stress to relax (or decay) toward a steady-state value. Additionally, the critical strain at which the shear stress exhibits an overshoot increases with $Wi$, but does not display a clear trend with respect to the elasticity number ($El$), as shown in Fig.~S4 of the supplementary materials. Despite these variations in the magnitude of the stress overshoot, the steady-state stress levels for both CTAB and CPyCl micellar solutions converge across different gap sizes (or elasticity number).}\\

\begin{figure}[hthp]
\includegraphics[width=0.9\textwidth]{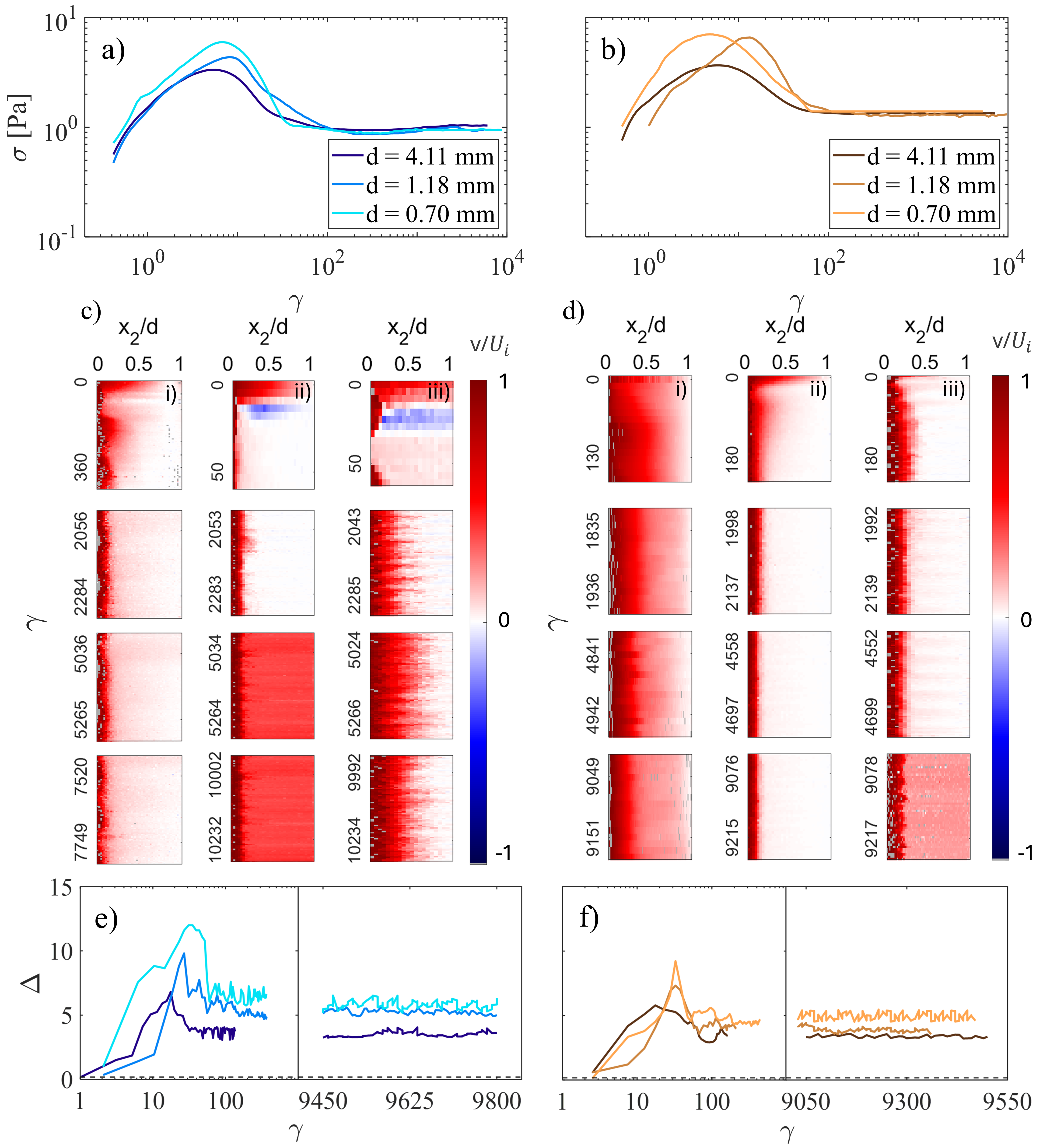}
\centering
\caption{Temporal evolution of shear stress for CTAB (a) and CPyCl (b) based wormlike micelles at Wi = 100. The 2D flow evolution at increasing $El$ value from left to right \AS{($El = 1.28\times10^5$ (i), $El = 1.55\times10^6$ (ii), and $El = 4.4\times10^6$ (iii) for CTAB (a) ($El = 1.3\times10^5$ (i), $El = 1.58\times10^6$ (ii), and $El = 4.48\times10^6$ (iii)) for CPyCl (d) systems. Note that due to slight differences in relaxation time and zero shear viscosity of CTAB and CPyCl systems, the fluid elasticities differ by about 1.8\%.} Temporal evolution \(\Delta\) for e) CTAB and f) CPyCl based systems at the TC cell gap sizes shown in (a) and (b).}
\label{fig: E_Wi100}
\end{figure}

Fig.~\ref{fig: E_Wi100}(c,d) present the spatio-temporal evolution of 2D velocity profiles for the two wormlike micellar systems, highlighting significant differences in the local rheology of each system. For both systems, and at the largest gap size (or weakest elasticity number; (i) column in Fig.~\ref{fig: E_Wi100}(c,d)), the initial transients do not show any signs of transient flow reversal. However, as the gap size decreases—or equivalently, as the elasticity number increases beyond a critical value of $El > 4.33\times10^5$, transient flow reversal (TFR) appears shortly after the stress decay (denoted by blue color in 2D profiles) for the CTAB system. In contrast, for the CPyCl-based micellar solution, this flow feature does not develop for any of the gap sizes tested, even when the elasticity number is as high as $ 4.47\times10^6 $. These results suggest two important aspects of shear banding flows. First, they confirm the existence of a critical elasticity number, beyond which transient flow reversal is observed, consistent with previous experimental observations~\cite{rassolov2020effects,mohammadigoushki2019transient} as well as predictions from the VCM model~\cite{zhou2014wormlike,mohammadigoushki2019transient}. In previous experiments~\cite{rassolov2020effects}, the elasticity number was varied by significantly altering surfactant and/or salt concentrations (in some cases, by an order of magnitude), potentially influencing other system properties, including transitions between concentration regimes, which may affect flow behavior. The present results affirm that the elasticity number is a crucial factor influencing the transient development of flow profiles in wormlike micelle systems. Second, surfactant chemistry plays a key role in shaping the evolution of flow profiles in shear-banding wormlike micelles, even under otherwise identical bulk rheological conditions. \\ 

Fig.~\ref{fig: E_Wi100}(c,d) present the spatio-temporal evolution of velocity profiles (referred to as 2D flow profiles from now on) at larger strain values, approaching the quasi-steady limit. As time progresses toward these quasi-steady conditions, the velocity profiles differ notably across varying the elasticity number and between the two wormlike micellar systems. At the smallest elasticity number (i in Fig.~\ref{fig: E_Wi100}(c,d)), the CTAB-based system exhibits a shear-banded flow, where distinct high and low shear bands appear. To aid interpretation of velocity maps, representative 1D velocity profiles for Wi = 100 are included in the Fig.~S5 of the supplementary materials. As the gap size decreases, the high-shear band remains visible, but the low-shear band becomes more mobile. Eventually, at $d$ = 1.18 mm (ii in Fig.~\ref{fig: E_Wi100}(c,d)), the low-shear band transitions to a plug-like behavior and shows significant wall slip at the outer cylinder. Surprisingly, at the smallest gap size (corresponding to the highest elasticity number), the quasi-steady shear-banding flow undergoes a dramatic change, with the high-shear band occupying a larger portion of the TC cell gap and exhibiting an unstable nature, marked by visible time-dependent fluctuations. The evolution of the quasi-steady flows in the CPyCl system differs from that of CTAB. At the lowest elasticity number, the high-shear band forms near the inner cylinder, while the low-shear band shows significant wall slip at the outer cylinder. As the gap size decreases (or elasticity number increases), the low-shear band becomes less mobile, and at higher fluid elasticities, it again shows wall slip. These observations highlight distinct differences in the flow behavior of the two systems, and we will revisit these features in a later section.\\
 
Fig.~\ref{fig: E_Wi100}(e,f) display the shear rate inhomogeneity $\Delta$ for both the CTAB and CPyCl-based systems at $Wi = 100$. For both systems, $\Delta$ initially increases and reaches a maximum value during the transient phase of the experiment. This maximum occurs at a strain value slightly beyond the point where the shear stress reaches its overshoot. Subsequently, $\Delta$ decays to a more steady value as strain increases. For conditions that lead to transient flow reversal (TFR), the intensity of TFR correlates with the magnitude of the $\Delta$ overshoot; stronger TFR is associated with higher $\Delta_{\text{max}}$ values. \AS{When comparing the two surfactant systems, CTAB exhibited higher $\Delta_{\text{max}}$ values than CPyCl at a given TC cell gap size.} Additionally, during the quasi-steady state, the $\Delta$ values increase with elasticity number for both systems, indicating that a higher elasticity number enhances flow inhomogeneity. Similar features (stress overshoot behavior, transient and quasi-steady flow) have been observed for these two wormlike micellar solutions at other imposed $Wi$ numbers within the stress plateau regime (e.g., $Wi$ = 200 and 400; see representative data in Fig.~S6-S7 in the supplementary materials).\\ 

Figure~(\ref{fig: KineticsE}) provides a comprehensive summary of the experimental results, highlighting key aspects of shear-banding flows, including transient flow reversal (TFR), the degree of flow inhomogeneity, and wall slip at both the inner and outer cylinders of the TC cell for the two wormlike micellar solutions across a wide range of Weissenberg numbers $Wi$ and fluid elasticities tested in this study. We begin by examining the impact of the imposed $Wi$ on the strength of the transient flow reversal across the range of gap sizes used in the experiments. Fig.~\ref{fig: KineticsE}(a) presents the ratio of the maximum transient reversal velocity observed within the gap (typically occurring following the stress overshoot) to the imposed velocity at the inner cylinder as a function of $Wi$ for the two wormlike micellar solutions at various gap sizes. In cases where TFR is observed, the velocity becomes negative, and in the absence of TFR, we denote the maximum recoil velocity as zero. For the CTAB/NaSal-based system (shown in Fig.~\ref{fig: KineticsE}(a,i)), TFR is not observed for larger gap sizes (4.11 mm and 2.23 mm) over the entire range of imposed $Wi$. As the gap size decreases (or elasticity number increases), TFR becomes apparent for $Wi >$ 50, and its intensity increases as the elasticity number rises. These results confirm the existence of a critical elasticity number threshold, beyond which transient flow reversal occurs. This finding is consistent with previous results on different micellar systems~\cite{rassolov2020effects} and aligns with predictions from the VCM model~\cite{zhou2014wormlike}. Moreover, the strength of the TFR exhibits a non-monotonic dependence on $Wi$. At low $Wi$, the TFR is most pronounced at the smallest gap size ($d$ = 0.70 mm). As $Wi$ increases, the TFR strength intensifies until reaching a critical $Wi= $ 100, after which the TFR weakens. In contrast, for the CPyCl/NaSal-based micellar solution (Fig.~\ref{fig: KineticsE}(a,ii)), no signs of TFR were observed across a broad range of elasticity number and imposed $Wi$. \\

Fig.~\ref{fig: KineticsE}(b) show the degree of flow inhomogeneity for both surfactant systems over a wide range of $Wi$ numbers. The degree of flow inhomogeneity is quantified using two parameters: $\Delta_{max}$, which represents the maximum extent of flow inhomogeneity observed during the process, and $\Delta_{ss}$, which characterizes the flow inhomogeneity under steady-state conditions. For micellar solution based on CTAB, $\Delta_{max}$ is an increasing function of $Wi$, and elasticity number. These results indicate that the strength of the TFR is directly proportional to $\Delta_{max}$. For the micellar system based on CPyCl, $\Delta_{max}$ exhibits a similar trend with respect to the Weissenberg number ($Wi$) and the TC cell gap size. It is noteworthy that $\Delta_{max}$ is consistently higher for the CTAB system compared to the CPyCl system at a fixed $Wi$. The primary axis of Fig.~\ref{fig: KineticsE}(b) also displays the $\Delta_{ss}$ values for the two micellar solutions. For each system, $\Delta_{ss}$ increases with increasing elasticity number. Overall, $\Delta_{ss}$ is comparable between the two systems.\\ 

Fig.~\ref{fig: KineticsE}(c) illustrates the impact of the elasticity number and the imposed $Wi$ on wall slip at the inner and outer cylinders of the TC cell for the two micellar solutions. For both systems and across various fluid elasticities, wall slip at the inner cylinder remains minimal. However, wall slip at the outer cylinder is more pronounced in certain conditions. At the smallest fluid elasticities, wall slip at the outer cylinder is minimal. As the gap size decreases (or elasticity number increases), wall slip at the outer cylinder becomes more significant, exhibiting a non-monotonic dependence on $Wi$. For $Wi$ values close to the onset of shear banding, wall slip is at its maximum. As $Wi$ increases further, wall slip decreases for both the CTAB and CPyCl-based systems. As the elasticity number decreases (or the TC cell gap size increases), the non-monotonic trend in wall slip with respect to $Wi$ persists, though it becomes weaker, and the maximum wall slip shifts to higher $Wi$ numbers. Prior studies have primarily evaluated wall slip at the inner cylinder of the TC cell and typically reported negligible wall slip at the outer cylinder~\cite{becu2004spatiotemporal, helgeson2009rheology, hu2005kinetics, mohammadigoushki2018creeping, yamamoto2009velocity, becu2007evidence}. The differences observed in wall slip at the outer cylinder between our experiments and most prior studies are likely due to differences in elasticity number. The elasticity number in prior studies was typically in the range of $\mathcal{O}(10^1 - 10^4)$~\cite{hu2005kinetics, miller2007transient, lettinga2009competition}, much lower than the values used in this study $\mathcal{O}(10^5 - 10^6)$, as well as in our earlier work~\cite{mohammadigoushki2019transient}. \HM{The impact of fluid elasticity on wall slip at the outer cylinder is not yet fully understood and remains to be systematically explored through theoretical modeling and experiments. }\\
\begin{figure}[hthp]
\includegraphics[width=1\textwidth]{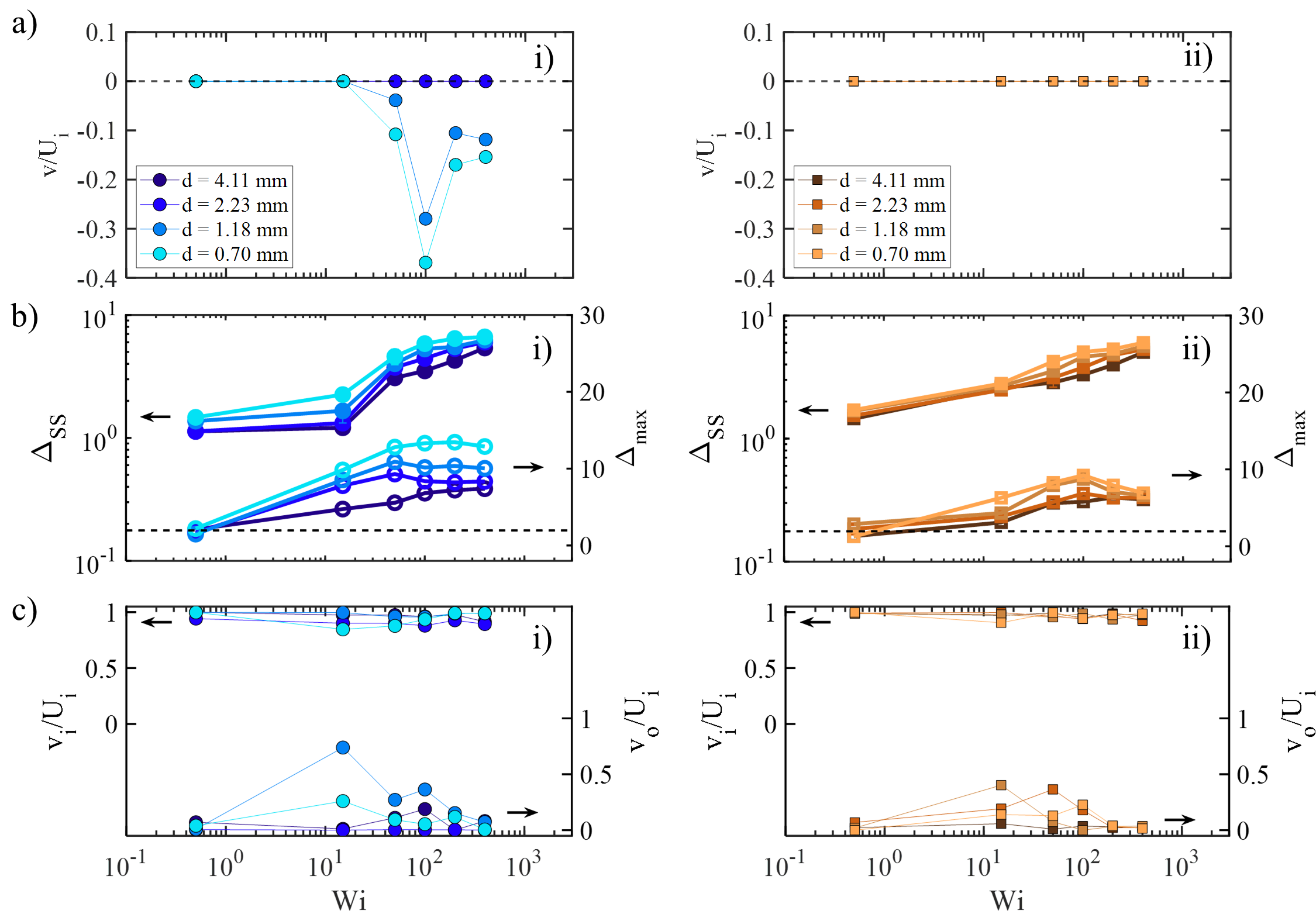}
\centering
\caption{Kinetics of shear banding flow formation at different $Wi$ numbers and fluid elasticities for i) CTAB and ii) CPyCl based systems. a) Normalized maximum transient fluid reversal velocity. b) flow inhomogeneity parameter $\Delta$ as a function of $Wi$,  where empty markers denote the maximum value observed during overshoot and the filled markers represent the quasi-steady state values. c) normalized wall slip velocity at the inner (top data) and outer (bottom data) cylinders as a function of $Wi$.}
\label{fig: KineticsE}
\end{figure}

\subsubsection{Effect of TC cell curvature:}
\begin{figure}[hthp]
\includegraphics[width=0.9\textwidth]{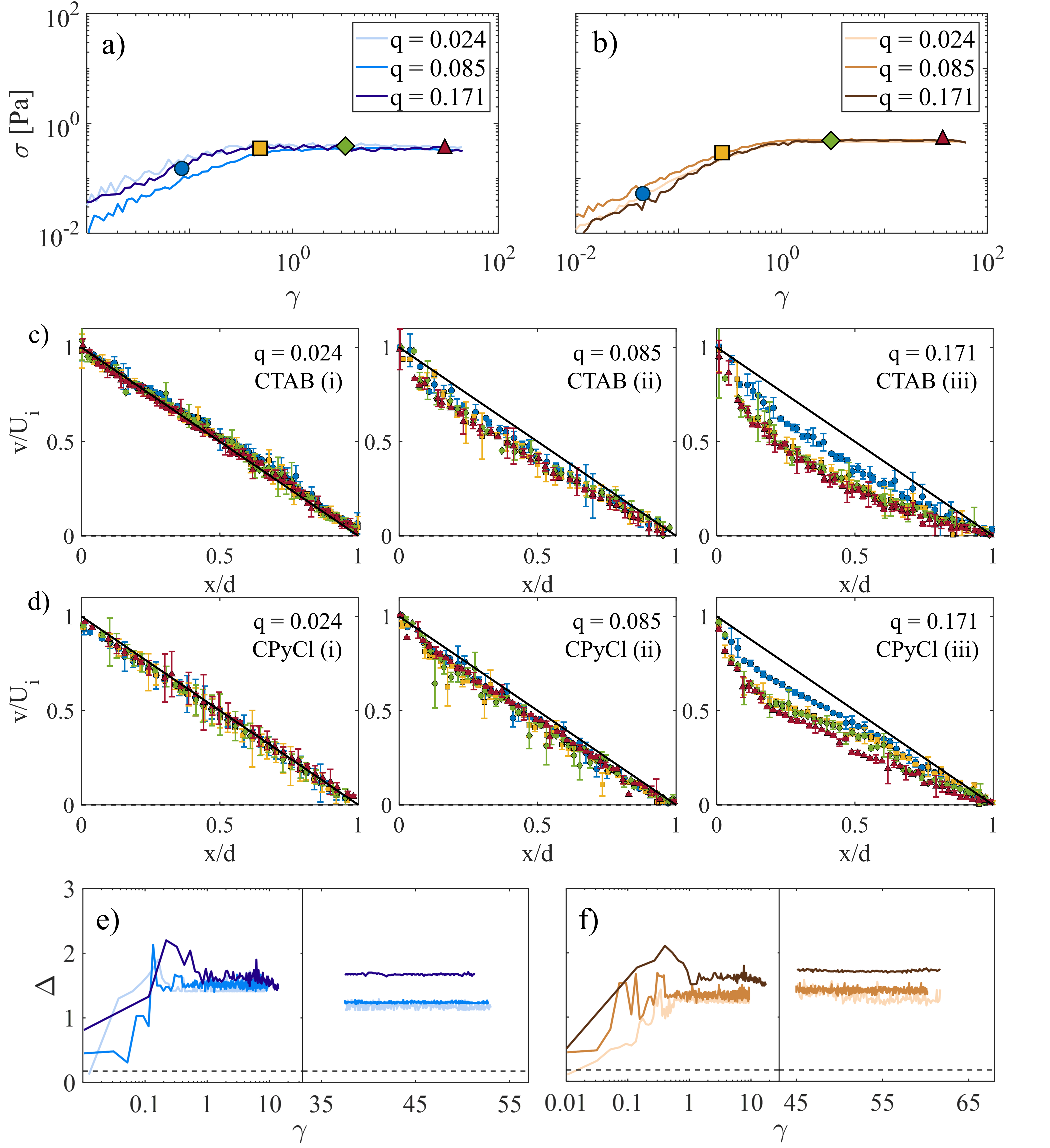}
\centering
\caption{Temporal evolution of shear stress for CTAB (a) and CPyCl (b) systems at $Wi$ = 0.5. The 1D velocity profiles are shown for various flow curvatures for CTAB (c) and CPyCl (d) systems. The symbols in (a,b) shear stress curves correspond to those plotted on the corresponding (c,d) velocity profiles. The temporal evolution of flow inhomogeneities for CTAB (e) and CPyCl (f) micellar systems at various flow curvatures shown in parts (a) and (b).}
\label{fig: q_Wi0.5}
\end{figure}

We now turn our attention to the impact of TC cell curvature on the transient evolution of flows toward quasi-steady shear banding flow formation. Bulk rheological measurements indicate that the TC cell curvature (within the range used in this study) does not significantly affect the quasi-steady flow curves and the shear stress plateau (see Fig.~S8 in the supplementary materials). As shown in Fig.~\ref{fig: q_Wi0.5}, at low imposed $Wi$ numbers below the onset of shear banding (e.g., $Wi$ = 0.5), the transient evolution of bulk shear stress for both CTAB and CPyCl solutions appears relatively smooth, without significant fluctuations. Although shear stresses are similar at different TC curvatures, the transient evolution of velocity profiles is different (see Fig.~\ref{fig: q_Wi0.5}(c,d)). As TC cell curvature increases, the deviation from Newtonian velocity behavior increases. The latter behavior is reflected in the shear rate inhomogeneity parameter $\Delta$, which is higher for larger TC cell curvatures. This behavior is expected because the TC cell curvature is expected to introduce a gradient in the shear stress, with shear stress being highest at the inner cylinder and lowest close to the stationary outer cylinder. Therefore, even a slight amount of shear thinning and stress inhomogeneity is expected to produce a curved velocity profile. Finally, no transient flow reversal is detected at this $Wi$ number for all curvature values. \\

At higher imposed $Wi$ numbers within the stress plateau, we observe notable changes in the transient stress response, which depend on the flow curvature of the TC cell. Fig.~\ref{fig: q_Wi100}(a,b) illustrate the temporal evolution of shear stress as a function of applied strain for CTAB and CPyCl solutions at $Wi$ = 100. As flow curvature increases, the transient stress overshoot becomes more pronounced for both CTAB and CPyCl solutions. This stress overshoot was measured across various flow curvatures and imposed $Wi$ numbers. Results (shown as Fig.~S9 in the supplementary materials) reveal that increasing flow curvature intensifies the stress overshoot. DRP fluid model simulations indicate that stress overshoot remains relatively unchanged for $q = 10^{-4} - 10^{-2}$. Note that the $q$ values explored in this study fall outside the range examined by previous DRP fluid model investigations~\cite{adams2011transient}. Similar to results on the effect of $El$, the critical strain at which the shear stress exhibits an overshoot increases with the $Wi$, but does not display a clear trend with respect to $q$, as shown in Fig.~S10 of the supplementary materials. The reason behind the observed enhancement of the stress overshoot with larger flow curvature for these wormlike micellar solutions remains an open question. In addition, the DRP fluid model predictions suggest that stress relaxes more rapidly in geometries with stronger stress gradients (i.e., larger $q$ values), which aligns with our experimental findings shown in Fig.~\ref{fig: q_Wi100}(a,b) and also for higher imposed $Wi$ numbers.\\

Fig.~\ref{fig: q_Wi100}(c,d) display the 2D velocity profiles across the gap of the TC cells, normalized by the inner cylinder velocity $U_i$, for both wormlike micelles based on CTAB and CPyCl. To aid interpretation of velocity maps, representative 1D velocity profiles for Wi = 100 are included in Fig.~S11 of the supplementary materials. At $Wi$ = 100, during the initial transients following the stress overshoot, TFR is not observed at low flow curvature ($q$ = 0.024) for either the CTAB or CPyCl systems (see the first column from left in Fig.~\ref{fig: q_Wi100}(c,d)). However, beyond a critical flow curvature ($q\geq$ 0.085), TFR is observed in the CTAB-based wormlike micellar solutions. As flow curvature increases, the intensity of the TFR strengthens for the CTAB system. In contrast, no TFR is observed for the CPyCl solution at any curvature values explored in this study, despite having the same bulk rheology as that of the CTAB/NaSal system. At higher $Wi$ numbers ($Wi = 200 $ or $Wi = 400$) and $q = 0.171$, the 2D velocity profiles of CPyCl solution show some signature of velocity recoil (see Fig.~S12-13 of the supplementary materials). Note that for $q = 0.171$ and $Wi = 200$ stress fully decays at $\gamma \approx 40$, while the fluctuations in velocity profile start from $\gamma \approx {160}$. Furthermore, this pattern persists even at much higher strain values. The typical TFR vanishes before stress reaches a quasi-steady level. These observations support the conclusion that the emerging patterns for CPyCl solution at higher $Wi$ and $q = 0.171$ are related to elastic instabilities or the slight wobbling of the inner cylinder rather than a transient flow reversal during stress decay period. On the other hand, the experimental results for CTAB during transient times are consistent with the predictions of the DRP fluid model. Adams et al.~\cite{adams2011transient} predicted that as flow curvature increases, the transient flow reversal becomes more pronounced. This was rationalized through stress gradients that are developed within the gap of the TC cell. As curvature increases, the stress gradient becomes more pronounced, with stress being higher close to the inner cylinder. This may result in the formation of high-velocity gradients (or shear bands) near the inner cylinder, potentially triggering transient flow reversal. \\ 

At higher strain values, the spatio-temporal evolution of the velocity profiles reveals intriguing dynamics as shown in Fig.~\ref{fig: q_Wi100}(c,d). In the CTAB system, under conditions of the lowest flow curvature, the expected shear-banded flow, characterized by well-defined high and low shear bands, is notably absent, which is unexpected. To investigate this phenomenon further, we conducted experiments at higher Weissenberg numbers $Wi$ = 200 and 400. Even at these elevated $Wi$ values and lowest $q$, shear-banded velocity profiles do not emerge at long strain values, despite the presence of a clear stress plateau in quasi-steady flow curve measurements. As the flow curvature increases to $q$ = 0.085, a well-defined shear-banded flow emerges in the CTAB system, accompanied by significant wall slip at the outer cylinder. However, upon further increasing the flow curvature to $q$ = 0.171, the wall slip at the outer cylinder vanishes, and a stable two-banded flow structure is recovered. To distinguish between shear-banded and non-banded velocity profiles, we analyze the one-dimensional velocity profiles during the quasi-steady state. Shear banding is characterized by two distinct regions with nearly constant shear rates separated by a narrow interface, whereas non-banded profiles typically exhibit a smoothly varying, curved velocity profile. Taking the first derivative of the velocity reveals this distinction more clearly: shear-banded profiles produce sharp, localized gradients, while non-banded profiles result in broader, more distributed gradients (e.g., Fig~S14 in the supplementary materials). In contrast, the CPyCl solution consistently exhibits a robust two-banded shear flow across all tested curvatures, with pronounced instabilities in the high shear band and persistent wall slip at the outer cylinder. As the flow curvature increases, the shear-banded structure in the CPyCl system remains largely unchanged, with only minimal variations in wall slip at the outer boundary. Interestingly, at the highest flow curvature, significant wall slip at the outer cylinder reemerges, suggesting a complex interplay between flow curvature, banding stability, and interfacial slip dynamics in wormlike micellar solutions.\\

\begin{figure}[hthp]
\includegraphics[width=0.9\textwidth]{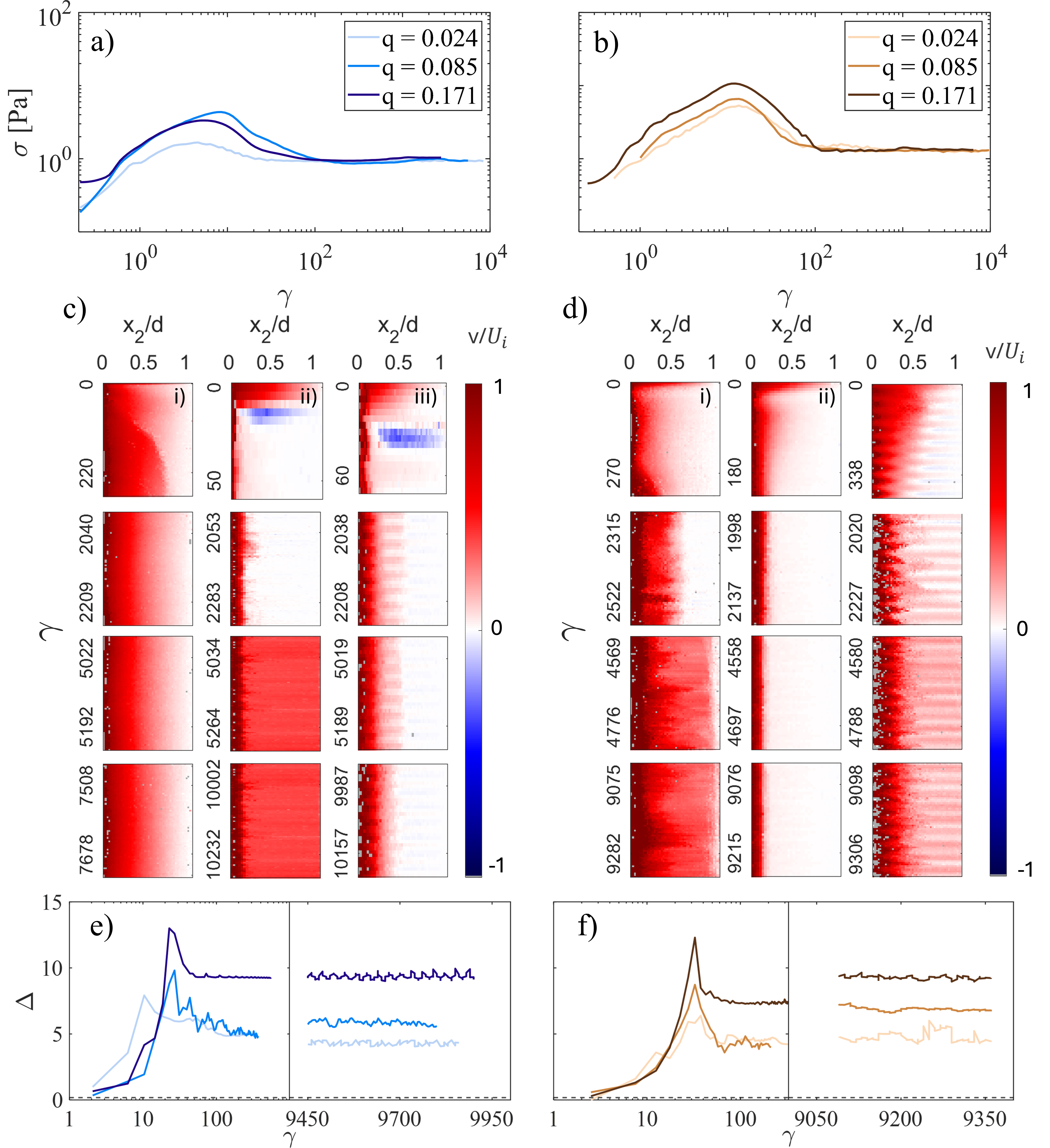}
\centering
\caption{Temporal evolution of shear stress for CTAB (a) and CPyCl (b) based micellar solutions at $Wi$ = 100. Spatio-temporal evolution of flow profiles in the wormlike micellar solutions based on CTAB (c) and CPyCl (d) at increasing $q$ value from left to right ($q = 0.024$ for (i), $q = 0.085$  for (ii), and $q = 0.171$ for (iii)). The temporal evolution of flow inhomogeneities for CTAB (e) and CPyCl (f) micellar systems at various flow curvatures shown in parts (a) and (b).}
\label{fig: q_Wi100}
\end{figure}

In Fig.~\ref{fig: q_Wi100}(e,f), the temporal evolution of flow inhomogeneities ($\Delta$) is presented for both systems at various TC flow curvatures. For both systems, the flow becomes most inhomogeneous (peak in $\Delta$) shortly after the stress overshoot. In the CTAB system, the transient overshoot in $\Delta$ is weakest at the smallest flow curvature; however, as the flow curvature increases, flow inhomogeneities are amplified during both the transient and quasi-steady states. A similar trend is reported in micellar solutions based on CPyCl/NaSal. These experimental observations align with predictions of the DRP model. The absence of TFR in the CPyCl solution at any curvature value further underscores the importance of surfactant chemistry in understanding the flow dynamics of wormlike micelle systems in curved geometries.\\

Figure~(\ref{fig: Kineticsq}) summarizes the impact of flow curvature on various aspects of the kinetics of shear banding flow formation in the wormlike micelles over a broad range of imposed $Wi$ numbers. In Fig.~\ref{fig: Kineticsq}(a), the normalized maximum transient flow reversal velocity is shown as a function of $Wi$ for the range of TC cell curvatures used in this study. At low flow curvatures, the transient flow reversal is not observed over the entire range of $Wi$ numbers. Note that a normalized velocity of zero in Fig.~\ref{fig: Kineticsq}(a) denotes no TFR. As flow curvature increases, the TFR is strengthened for the CTAB based system, while for the CPyCl system, regardless of the TC cell curvature, we do not observe TFR. In addition, the impact of the flow curvature on transient flow reversal observed for CTAB/NaSal flow is highest at an intermediate $Wi$ number ($Wi$ = 100). As noted before, the observed behavior for wormlike micellar solution based on CTAB, where TFR strengthens as $q$ increases, is consistent with theoretical predictions of the DRP fluid model~\cite{adams2011transient}. Larger values of $q$ correspond to greater stress gradients across the TC gap. These stress gradients could destabilize homogeneous flow, and may enhance the formation of shear bands, particularly during transient times. This leads to more pronounced velocity differences between bands during startup flows, and that may, in turn, lead to a more likely formation of the transient flow reversal. \\

Furthermore, Fig.~\ref{fig: Kineticsq}(b) presents the shear rate inhomogeneity parameter, comparing the maximum $\Delta_{\text{max}}$ and steady-state $\Delta_{\text{SS}}$ values for both CTAB and CPyCl systems as a function of imposed $Wi$ across the flow curvature range examined in this study. For both systems, the degree of flow inhomogeneities (both for transient and quasi-steady conditions) increase as the flow curvature increases. For CTAB samples, $\Delta_{\text{max}}$ is consistently higher than those reported for CPyCl, otherwise at similar rheological and flow curvature features. It is also worth noting that at $Wi$ = 100, where the strongest TFR was observed $\Delta_{\text{max}}$ reaches an apparent maximum before slightly leveling down at higher $Wi$ numbers. This suggests a strong correlation between the occurrence of TFR and the observed maxima in $\Delta_{\text{max}}$. These observations on the micellar solutions based on CTAB align with the findings of Adams and coworkers, where $\Delta_{\text{max}}$ was shown to increase with $q$, showcasing its sensitivity to both the imposed flow conditions and the rheological response of the system.\\ 

Fig.~\ref{fig: Kineticsq}(c) illustrates the wall slip behavior at both the inner and outer cylinders for the two micellar systems. The normalized velocity at the inner cylinder, $v_i/U_i$, remains close to unity for low $Wi$ across all tested $q$-values, with only minor variations as $Wi$ increases. Notably, the lowest normalized velocity occurs within the range $10 \leq Wi \leq 100$ for both systems, but no clear trend emerges beyond this observation. In contrast, the slip velocity at the outer cylinder, $ v_o/U_i$, starts near zero at low $Wi$ and progressively increases with $Wi$, reaching its peak at approximately $Wi$ = 15. This trend is evident in both systems but is more pronounced in the CTAB solution, indicating a greater sensitivity to flow curvature effects at the outer cylinder. Interestingly, the relationship between outer wall slip and TC cell curvature is non-monotonic: at the lowest curvature ($q$ = 0.024), wall slip is negligible, then it increases to a maximum at $q$ = 0.085, before decreasing again at higher curvatures. Additionally, the CPyCl system consistently exhibits weaker outer wall slip compared to the CTAB-based system under otherwise identical flow conditions. \\

\begin{figure}[hthp]
\includegraphics[width=1\textwidth]{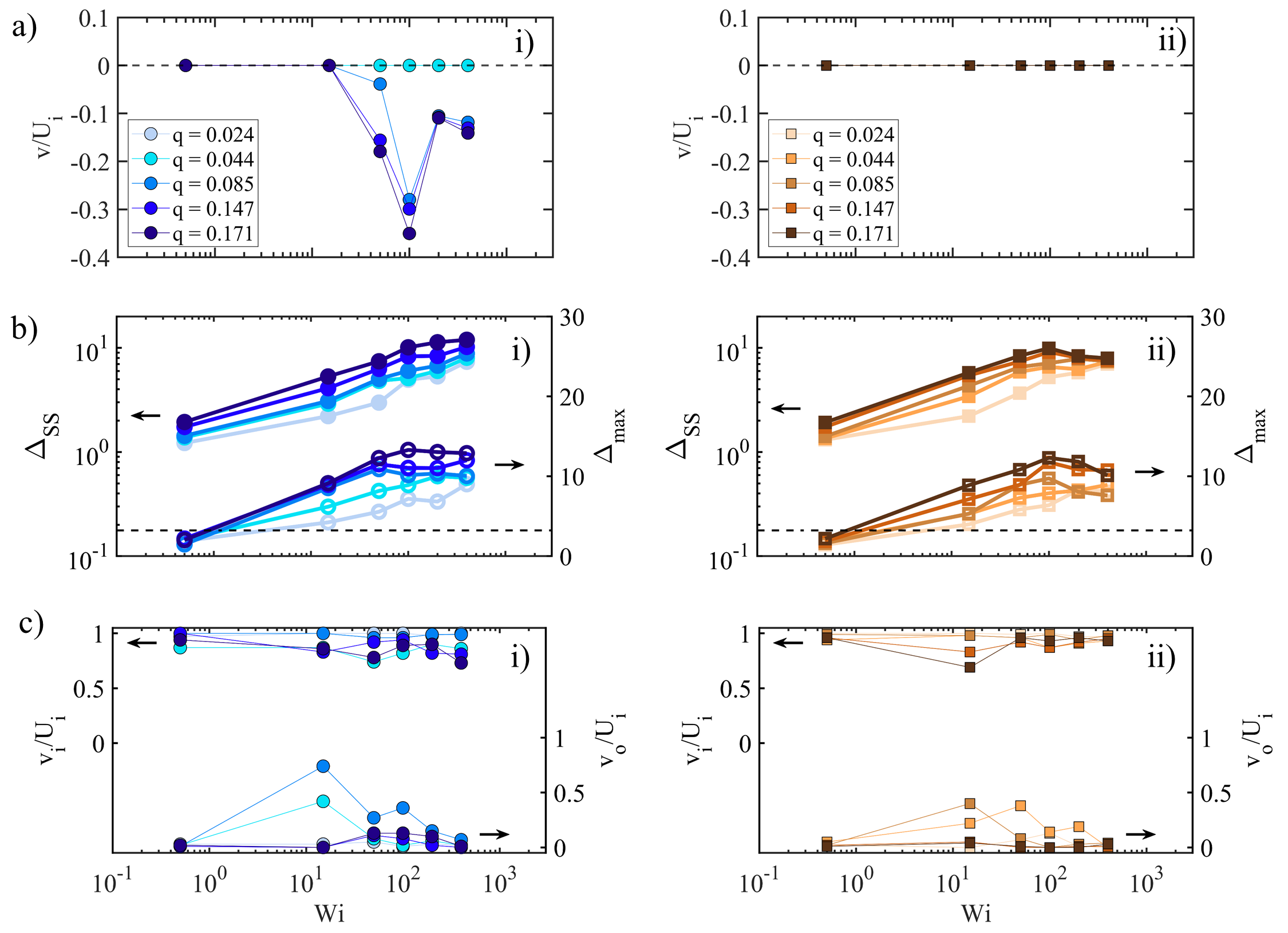}
\centering
\caption{Kinetics of shear banding flow formation at different $Wi$ numbers and flow curvatures for i) CTAB and ii) CPyCl based systems. a) Normalized maximum of the fluid velocity during transient flow reversal. b) flow inhomogeneity parameter $\Delta$ as a function of $Wi$,  where empty markers denote the maximum value observed during overshoot and the filled markers represent the quasi-steady state values. c) normalized wall slip velocity at the inner (top data) and outer (bottom data) cylinders as a function of $Wi$.}
\label{fig: Kineticsq}
\end{figure}

Finally, we revisit the observed differences in shear banding flow kinetics between CTAB/NaSal and CPyCl/NaSal wormlike micellar solutions. Despite their similar bulk rheological properties, these systems exhibit distinct flow evolution dynamics. Both solutions are formulated at a surfactant-to-salt concentration ratio below the viscosity maximum, where a predominantly linear micellar structure is expected. One key microstructural parameter influencing flow behavior is the average contour length of the micelles, \( \langle L \rangle \), which may differ between these systems. The linear viscoelastic response suggests that both micellar solutions operate in the fast-breaking regime, where the micellar breakage time (\(\tau_{br}\)) is much shorter than the reptation time (\(\tau_{rep}\)), i.e., \( \tau_{br} \ll \tau_{rep} \). The breakage time is estimated as \( \tau_{br} \approx 1/\omega_{\text{min}} \), where \( \omega_{\text{min}} \) corresponds to the angular frequency at the local minimum of the loss modulus. Additionally, the longest relaxation time in the fast-breaking limit is given by \( \lambda = \sqrt{\tau_{rep} \tau_{br}} \). By leveraging literature values for the micellar persistence length $l_p$ ~\cite{oelschlaeger2009linear, galvan2008diffusing}, we estimate the micellar entanglement length $l_e$ using the relation $G_0 = \frac{k_B T}{l_e^{9/5} l_p^{6/5}} $. The corresponding average micellar contour length is then obtained as \( \langle L \rangle = Z l_e \), where \( Z \) is the number of entanglements per micelle. Our calculations yield an estimated micelle contour length of approximately 20 $\mu$m for both solutions. Since this value significantly exceeds the persistence length, the micelles are expected to be flexible. Given the comparable bulk rheology and microstructural characteristics of these two shear-banding wormlike micellar solutions, one would not anticipate significant differences in their local flow responses. This suggests that the observed disparities in shear banding dynamics may arise from more subtle factors, such as different chemistry of surfactants.\\

From a molecular standpoint, the fundamental differences between CTAB and CPyCl surfactant molecules arise from the size of their head groups and the nature of their counterions (Cl$^-$ vs. Br$^-$). The CPyCl head group is larger by approximately 40$\%$ than that of CTAB, which amplifies electrostatic repulsions between adjacent CPyCl surfactant molecules~\cite{israelachvili2011intermolecular,lutz2016viscoelasticity,lutz2017intermicellar}. As a result, at a fixed sodium salicylate (NaSal) concentration, CPyCl micelles may experience incomplete charge screening, leading to weaker intermolecular interactions and, consequently, lower bulk rheological properties compared to CTAB/NaSal~\cite{lutz2016viscoelasticity, lutz2017intermicellar}. This hypothesis is further supported by our formulation experiments, which demonstrated that achieving comparable bulk rheological properties between these two micellar systems necessitates a higher concentration of CPyCl and an increased NaSal-to-CPyCl ratio. Furthermore, literature reports indicate that Cl$^-$ counterions exhibit a higher dissociation rate than Br$^-$~\cite{bales2002characterization,buckingham1993effect}, suggesting that CPyCl/NaSal systems may have a greater proportion of free counterions in solution, thereby further weakening electrostatic screening. \\

{Previous studies have shown that the packing density of surfactants within micelles plays a central role in their resistance to deformation~\cite{dhakal2016uniaxial}. For example, molecular dynamics (MD) simulations have demonstrated that reducing electrostatic repulsion between headgroups leads to tighter and more cohesive molecular assemblies with enhanced resistance to extensional flow~\cite{dhakal2016uniaxial}. While our study focuses on shear deformation, the initial elastic stretching of micelles, and their resistance to deformation, should similarly depend on surfactant packing density. This linear packing density, defined as the number of surfactant molecules per unit length of micelle, can be estimated as $\rho_L = 2\pi R/a_0$, where $R$ is the micelle radius and $a_0$ is the effective area per headgroup. Based on literature data~\cite{schubert2003microstructure,lam2019structural}, we estimate $\rho_L \approx 23$ molecules/nm for CTAB and $\rho_L \approx 15$ molecules/nm for CPyCl, suggesting that CTAB micelles are more tightly packed by surfactants. Following this analysis, and in line with previous MD simulation results~\cite{dhakal2016uniaxial}, we hypothesize that the lower packing density of CPyCl micelles leads to weaker intermolecular cohesion, resulting in a less elastic and less resistant micellar network under transient flow. In contrast, the tighter packing in CTAB-based micelles likely contributes to a more cohesive assembly that resists transient deformation more effectively, giving rise to a stronger transient elastic recoil upon perturbation. }\\

{Several theoretical models have been developed to describe the flow and deformation of wormlike micelles, with the Cates model being one of the earliest to incorporate the role of micellar chemistry~\cite{cates1987reptation}. In the linear viscoelastic regime, the model predicts that the average micelle length depends on the micellar scission energy $E_{scission}$, which is also related to the micelle breakage time in the fast-breaking limit. According to the theory of Cates and Granek~\cite{granek1992stress}, the scission energy is related to the viscoelastic moduli $G'$ and $G"$, which provides a link between molecular interactions and bulk rheological properties. In the nonlinear viscoelastic regime, this framework has been extended by coupling micellar breakage and recombination kinetics with the tube model to explain complex flow behaviors such as shear banding~\cite{spenley1993nonlinear}. In our experiments, both CTAB/NaSal and CPyCl/NaSal solutions exhibit similar linear and non-linear rheological behavior, including comparable viscoelastic moduli and flow curves. Despite this similarity, their transient responses to the onset of flow differ significantly. This indicates that additional chemistry-dependent factors, such as differences in headgroup interactions, micellar packing density, or surfactant-counterion binding, may influence micellar stability and dynamics in ways not captured by standard continuum models. In addition, under strong deformation, the rate of micellar breakage and reformation may deviate from the simplified, flow-independent kinetics assumed in the original Cates model~\cite{spenley1993nonlinear}. In shear-banding wormlike micelles, flow can induce micellar breakage or lead to concentration coupling, both of which can modify micellar dynamics in a chemistry-dependent manner. That is, the effect of flow on breakage and reformation rates may vary depending on surfactant molecular structure, counterion binding, and other chemical factors. These complex, flow-chemistry interactions are not currently captured in conventional models and merit further investigation. On the other hand, a parallel can be drawn with recent studies on polymer solutions, where polymer-solvent interactions and chemistry have been shown to affect flow behavior and have been incorporated into continuum models through an osmotic pressure term~\cite{cromer2014study,burroughs2021flow,burroughs2023flow}. Similarly, future models for wormlike micelles could be improved by explicitly including the effects of surfactant chemistry and counterion binding, to more accurately predict chemically-specific transient responses under flow.} Finally, molecular and/or coarse-grained dynamic simulations~\cite{castillo2011rheology,padding2008dynamics,boek2007flow,dhakal2016uniaxial}, capable of capturing both molecular-scale interactions and large-scale flow phenomena, present a powerful tool for testing and explaining the experimental findings of this study.\\

\section{Conclusions}
In summary, we have investigated the nonlinear local flow response of two shear-banding wormlike micellar solutions following the inception of a step shear across a broad range of imposed $Wi$, elasticity number, and flow curvature. The two micellar solutions, formulated with CTAB/NaSal and CPyCl/NaSal, were designed to exhibit comparable bulk rheological and microstructural properties. Our key findings are as follows:

\begin{itemize}
\item Experimental results reveal that as the elasticity number increases, both wormlike micellar solutions exhibit a pronounced stress overshoot, with the magnitude of the overshoot intensifying with increasing elasticity. Additionally, at low fluid elasticities (or large TC cell gap sizes), the initial transient flow exhibits weak heterogeneity, with a high shear band forming near the inner wall. However, the heterogeneities (characterized by the parameter $\Delta$) are not pronounced enough to induce transient flow reversal following stress overshoot. Beyond a critical elasticity number threshold of $El>1.58\times10^6$, the transient flow reversal is observed following stress overshoot. This flow feature becomes more pronounced in the CTAB/NaSal system as the elasticity number increases, consistent with our previous measurements~\cite{rassolov2020effects} and existing predictions of the VCM fluid model~\cite{zhou2012multiple}. Interestingly, even at the highest fluid elasticities ($El = 4.47\times10^6$), transient flow reversal is absent in the CPyCl/NaSal system, highlighting a fundamental difference in their nonlinear flow response.

\item Our results further highlight the significant role of flow curvature in the kinetics of shear banding flow formation in wormlike micellar solutions. Specifically, increasing flow curvature amplifies the stress overshoot, making it more pronounced. The local flow response is also strongly dependent on flow curvature. At low flow curvatures, the transient response remains weakly heterogeneous, characterized by the absence of transient flow reversal and quasi-steady shear-banded flows at higher strains. However, as flow curvature increases, stronger stress gradients develop in the TC gap, leading to more pronounced transient flow reversal, consistent with predictions of the DRP fluid model~\cite{adams2011transient}. Notably, regardless of flow curvature, transient flow reversal was not observed in the CPyCl-based micellar solution.
\end{itemize}

Our hypothesis is that the observed differences between CTAB and CPyCl-based micelles are primarily due to the stronger electrostatic repulsions present in CPyCl surfactant solutions. While prior models have explored the effects of elasticity number and flow curvature on the behavior of wormlike micelles, a comprehensive framework that fully incorporates surfactant chemistry into the non-linear flow dynamics of shear-banding wormlike micelles is still lacking. We hope this study will pave the way for the development of next-generation, chemistry-informed rheological models for the flows of viscoelastic wormlike micellar solutions.\\

\section{Supplementary Material}
See the supplementary material for details on the bulk and local rheology data in the two wormlike micellar solutions.

\section{Acknowledgments}
We gratefully acknowledge the National Science Foundation through the CBET CAREER award 1942150 for funding this study.

\section{Author Declarations}
\subsection{Conflict of Interest}
The authors have no conflicts to disclose.

\bibliographystyle{unsrt}
\bibliography{BIB.bib}

\begin{thebibliography}{10}

\bibitem{cates2006rheology}
Michael~E Cates and Suzanne~M Fielding.
\newblock Rheology of giant micelles.
\newblock {\em Advances in Physics}, 55(7-8):799--879, 2006.

\bibitem{dreiss2007wormlike}
Cecile~A Dreiss.
\newblock Wormlike micelles: where do we stand? recent developments, linear rheology and scattering techniques.
\newblock {\em Soft Matter}, 3(8):956--970, 2007.

\bibitem{chu2013smart}
Zonglin Chu, C{\'e}cile~A Dreiss, and Yujun Feng.
\newblock Smart wormlike micelles.
\newblock {\em Chemical Society Reviews}, 42(17):7174--7203, 2013.

\bibitem{israelachvili2011intermolecular}
Jacob~N Israelachvili.
\newblock {\em Intermolecular and surface forces}.
\newblock Academic press, 2011.

\bibitem{yang2002viscoelastic}
Jiang Yang.
\newblock Viscoelastic wormlike micelles and their applications.
\newblock {\em Current opinion in colloid \& interface science}, 7(5-6):276--281, 2002.

\bibitem{rothstein2020complex}
Jonathan~P Rothstein and Hadi Mohammadigoushki.
\newblock Complex flows of viscoelastic wormlike micelle solutions.
\newblock {\em Journal of Non-Newtonian Fluid Mechanics}, 285:104382, 2020.

\bibitem{olmsted2008perspectives}
Peter~D Olmsted.
\newblock Perspectives on shear banding in complex fluids.
\newblock {\em Rheologica Acta}, 47(3):283--300, 2008.

\bibitem{divoux2016shear}
Thibaut Divoux, Marc~A Fardin, Sebastien Manneville, and Sandra Lerouge.
\newblock Shear banding of complex fluids.
\newblock {\em Annual Review of Fluid Mechanics}, 48(1):81--103, 2016.

\bibitem{rehage1991viscoelastic}
H~Rehage and H~Hoffmann.
\newblock Viscoelastic surfactant solutions: model systems for rheological research.
\newblock {\em Molecular Physics}, 74(5):933--973, 1991.

\bibitem{cates2008shearing}
ME~Cates, SM~Fielding, D~Marenduzzo, E~Orlandini, and JM~Yeomans.
\newblock Shearing active gels close to the isotropic-nematic transition.
\newblock {\em Physical review letters}, 101(6):068102, 2008.

\bibitem{berret2005}
J.~F. Berret.
\newblock Equilibrium properties and shear-banding transition.
\newblock In {\em Molecular Gels}. Springer, Dordrecht, 2005.

\bibitem{rassolov2022kinetics}
Peter Rassolov, Alfredo Scigliani, and Hadi Mohammadigoushki.
\newblock Kinetics of shear banding flow formation in linear and branched wormlike micelles.
\newblock {\em Soft Matter}, 2022.

\bibitem{lerouge2010shear}
Sandra Lerouge and Jean-Fran{\c{c}}ois Berret.
\newblock Shear-induced transitions and instabilities in surfactant wormlike micelles.
\newblock {\em Polymer Characterization: Rheology, Laser Interferometry, Electrooptics}, pages 1--71, 2010.

\bibitem{berret1997inhomogeneous}
Jean-Fran{\c{c}}ois Berret, Gr{\'e}goire Porte, and Jean-Paul Decruppe.
\newblock Inhomogeneous shear flows of wormlike micelles: ma master dynamic phase diagram.
\newblock {\em Physical Review E}, 55(2):1668, 1997.

\bibitem{boukany2008use}
Pouyan~E Boukany and Shi-Qing Wang.
\newblock Use of particle-tracking velocimetry and flow birefringence to study nonlinear flow behavior of entangled wormlike micellar solution: From wall slip, bulk disentanglement to chain scission.
\newblock {\em Macromolecules}, 41(4):1455--1464, 2008.

\bibitem{britton1997two}
Melanie~M Britton and Paul~T Callaghan.
\newblock Two-phase shear band structures at uniform stress.
\newblock {\em Physical review letters}, 78(26):4930, 1997.

\bibitem{casanellas2015spatiotemporal}
Laura Casanellas, Christopher~J Dimitriou, Thomas~J Ober, and Gareth~H McKinley.
\newblock Spatiotemporal dynamics of multiple shear-banding events for viscoelastic micellar fluids in cone-plate shearing flows.
\newblock {\em Journal of Non-Newtonian Fluid Mechanics}, 222:234--247, 2015.

\bibitem{ravindranath2008banding}
Sham Ravindranath, Shi-Qing Wang, Michael Olechnowicz, and Roderic~P Quirk.
\newblock Banding in simple steady shear of entangled polymer solutions.
\newblock {\em Macromolecules}, 41(7):2663--2670, 2008.

\bibitem{britton1999transition}
MM~Britton, RW~Mair, RK~Lambert, and PT~Callaghan.
\newblock Transition to shear banding in pipe and couette flow of wormlike micellar solutions.
\newblock {\em Journal of Rheology}, 43(4):897--909, 1999.

\bibitem{mair1996observation}
RW~Mair and PT~Callaghan.
\newblock Observation of shear banding in worm-like micelles by nmr velocity imaging.
\newblock {\em Europhysics letters}, 36(9):719, 1996.

\bibitem{mair1997shear}
RW~Mair and PT~Callaghan.
\newblock Shear flow of wormlike micelles in pipe and cylindrical couette geometries as studied by nuclear magnetic resonance microscopy.
\newblock {\em Journal of Rheology}, 41(4):901--924, 1997.

\bibitem{brown2011changing}
Jennifer~R Brown and Paul~T Callaghan.
\newblock Changing micellar order, lever rule behavior and spatio-temporal dynamics in shear-banding at the onset of the stress plateau.
\newblock {\em Soft Matter}, 7(21):10472--10482, 2011.

\bibitem{hu2010steady}
Y~Thomas Hu.
\newblock Steady-state shear banding in entangled polymers?
\newblock {\em Journal of Rheology}, 54(6):1307--1323, 2010.

\bibitem{fardin2012instabilities}
M~A Fardin and S~Lerouge.
\newblock Instabilities in wormlike micelle systems: From shear-banding to elastic turbulence.
\newblock {\em The European Physicaljournal E}, 35:1--29, 2012.

\bibitem{fardin2012shear}
MA~Fardin, T~Divoux, MA~Guedeau-Boudeville, I~Buchet-Maulien, Julien Browaeys, GH~McKinley, S~Manneville, and S~Lerouge.
\newblock Shear-banding in surfactant wormlike micelles: Elastic instabilities and wall slip.
\newblock {\em Soft Matter}, 8(8):2535--2553, 2012.

\bibitem{mohammadigoushki2017inertio}
Hadi Mohammadigoushki and Susan~J Muller.
\newblock Inertio-elastic instability in taylor-couette flow of a model wormlike micellar system.
\newblock {\em Journal of Rheology}, 61(4):683--696, 2017.

\bibitem{mohammadigoushki2016flow}
Hadi Mohammadigoushki and Susan~J Muller.
\newblock A flow visualization and superposition rheology study of shear-banding wormlike micelle solutions.
\newblock {\em Soft matter}, 12(4):1051--1061, 2016.

\bibitem{helgeson2009rheology}
Matthew~E Helgeson, Paula~A Vasquez, Eric~W Kaler, and Norman~J Wagner.
\newblock Rheology and spatially resolved structure of cetyltrimethylammonium bromide wormlike micelles through the shear banding transition.
\newblock {\em Journal of Rheology}, 53(3):727--756, 2009.

\bibitem{salmon2003velocity}
J.~Salmon, A.~Colin, S.~Manneville, and F.~Molino.
\newblock Velocity profiles in shear-banding wormlike micelles.
\newblock {\em Physical Review Letters}, 90:228303, 2003.

\bibitem{lettinga2009competition}
M.~P. Lettinga and S.~Manneville.
\newblock Competition between shear banding and wall slip in wormlike micelles.
\newblock {\em Physical Review Letters}, 103:248302, 2009.

\bibitem{becu2004spatiotemporal}
Lydiane B{\'e}cu, S{\'e}bastien Manneville, and Annie Colin.
\newblock Spatiotemporal dynamics of wormlike micelles under shear.
\newblock {\em Physical review letters}, 93(1):018301, 2004.

\bibitem{masselon2010influence}
Chlo{\'e} Masselon, Annie Colin, and Peter~D Olmsted.
\newblock Influence of boundary conditions and confinement on nonlocal effects in flows of wormlike micellar systems.
\newblock {\em Physical Review E—Statistical, Nonlinear, and Soft Matter Physics}, 81(2):021502, 2010.

\bibitem{miller2007transient}
E.~Miller and J.~P. Rothstein.
\newblock Transient evolution of shear-banding wormlike micellar solutions.
\newblock {\em Journal of Non-Newtonian Fluid Mechanics}, 143:22--37, 2007.

\bibitem{hu2005kinetics}
Y~Thomas Hu and A~Lips.
\newblock Kinetics and mechanism of shear banding in an entangled micellar solution.
\newblock {\em Journal of Rheology}, 49(5):1001--1027, 2005.

\bibitem{mohammadigoushki2019transient}
Hadi Mohammadigoushki, Alireza Dalili, Lin Zhou, and Pamela Cook.
\newblock Transient evolution of flow profiles in a shear banding wormlike micellar solution: Experimental results and a comparison with the vcm model.
\newblock {\em Soft Matter}, 15(27):5483--5494, 2019.

\bibitem{gurnon2014spatially}
A~Kate Gurnon, Carlos Lopez-Barron, Matthew~J Wasbrough, Lionel Porcar, and Norman~J Wagner.
\newblock Spatially resolved concentration and segmental flow alignment in a shear-banding solution of polymer-like micelles.
\newblock {\em ACS Macro Letters}, 3(3):276--280, 2014.

\bibitem{feindel2010anomalous}
Kirk~W Feindel and Paul~T Callaghan.
\newblock Anomalous shear banding: multidimensional dynamics under fluctuating slip conditions.
\newblock {\em Rheologica acta}, 49:1003--1013, 2010.

\bibitem{al2018rheo}
Rehab~N Al-kaby, Jayesha~S Jayaratne, Timothy~I Brox, Sarah~L Codd, Joseph~D Seymour, and Jennifer~R Brown.
\newblock Rheo-nmr of transient and steady state shear banding under shear startup.
\newblock {\em Journal of Rheology}, 62(5):1125--1134, 2018.

\bibitem{kim2000effects}
Won-Jong Kim and Seung-Man Yang.
\newblock Effects of sodium salicylate on the microstructure of an aqueous micellar solution and its rheological responses.
\newblock {\em Journal of colloid and interface science}, 232(2):225--234, 2000.

\bibitem{osaki1979flow}
Kunihiro Osaki, Nobuo Bessho, Tetsuya Kojimoto, and Michio Kurata.
\newblock Flow birefringence of polymer solutions in time-dependent field.
\newblock {\em Journal of Rheology}, 23(4):457--475, 1979.

\bibitem{manneville2008recent}
S{\'e}bastien Manneville.
\newblock Recent experimental probes of shear banding.
\newblock {\em Rheologica Acta}, 47(3):301--318, 2008.

\bibitem{rassolov2020effects}
Peter Rassolov and Hadi Mohammadigoushki.
\newblock Effects of elasticity and flow ramp up on kinetics of shear banding flow formation in wormlike micellar fluids.
\newblock {\em Journal of Rheology}, 64(5):1161--1177, 2020.

\bibitem{rassolov2023role}
Peter Rassolov and Hadi Mohammadigoushki.
\newblock Role of micellar entanglements on kinetics of shear banding flow formation.
\newblock {\em Journal of Rheology}, 67(1):169--181, 2023.

\bibitem{mccauley2024heterogeneity}
Patrick~J McCauley, Satish Kumar, and Michelle~A Calabrese.
\newblock Heterogeneity-induced retraction in viscoelastic fluids following cessation of flow.
\newblock {\em Soft Matter}, 20(23):4567--4582, 2024.

\bibitem{mccauley2023evolution}
Patrick~J McCauley, Christine Huang, Lionel Porcar, Satish Kumar, and Michelle~A Calabrese.
\newblock Evolution of flow reversal and flow heterogeneities in high elasticity wormlike micelles (wlms) with a yield stress.
\newblock {\em Journal of Rheology}, 67(3):661--681, 2023.

\bibitem{vasquez2007network}
Paula~A Vasquez, Gareth~H McKinley, and L~Pamela Cook.
\newblock A network scission model for wormlike micellar solutions: I. model formulation and viscometric flow predictions.
\newblock {\em Journal of non-newtonian fluid mechanics}, 144(2-3):122--139, 2007.

\bibitem{zhou2014wormlike}
Lin Zhou, Gareth~H McKinley, and L~Pamela Cook.
\newblock Wormlike micellar solutions: Iii. vcm model predictions in steady and transient shearing flows.
\newblock {\em Journal of Non-Newtonian Fluid Mechanics}, 211:70--83, 2014.

\bibitem{pipe2010wormlike}
CJ~Pipe, NJ~Kim, PA~Vasquez, LP~Cook, and GH~McKinley.
\newblock Wormlike micellar solutions: Ii. comparison between experimental data and scission model predictions.
\newblock {\em Journal of Rheology}, 54(4):881--913, 2010.

\bibitem{germann2014investigation}
N~Germann, LP~Cook, and AN~Beris.
\newblock Investigation of the inhomogeneous shear flow of a wormlike micellar solution using a thermodynamically consistent model.
\newblock {\em Journal of Non-Newtonian Fluid Mechanics}, 207:21--31, 2014.

\bibitem{germann2013nonequilibrium}
Natalie Germann, LP~Cook, and Antony~N Beris.
\newblock Nonequilibrium thermodynamic modeling of the structure and rheology of concentrated wormlike micellar solutions.
\newblock {\em Journal of Non-Newtonian Fluid Mechanics}, 196:51--57, 2013.

\bibitem{bautista2002irreversible}
F~Bautista, JFA Soltero, ER~Macias, JE~Puig, and O~Manero.
\newblock Irreversible thermodynamics approach and modeling of shear-banding flow of wormlike micelles.
\newblock {\em The journal of Physical Chemistry B}, 106(50):13018--13026, 2002.

\bibitem{zhou2012multiple}
Lin Zhou, L~Pamela Cook, and Gareth~H McKinley.
\newblock Multiple shear-banding transitions for a model of wormlike micellar solutions.
\newblock {\em SIAMjournal on Applied Mathematics}, 72(4):1192--1212, 2012.

\bibitem{adams2011transient}
JM~Adams, Suzanne~M Fielding, and Peter~D Olmsted.
\newblock Transient shear banding in entangled polymers: A study using the rolie-poly model.
\newblock {\em Journal of Rheology}, 55(5):1007--1032, 2011.

\bibitem{fielding2007complex}
Suzanne~M Fielding.
\newblock Complex dynamics of shear banded flows.
\newblock {\em Soft Matter}, 3(10):1262--1279, 2007.

\bibitem{adams2008interplay}
JM~Adams, SM~Fielding, and PD~Olmsted.
\newblock The interplay between boundary conditions and flow geometries in shear banding: Hysteresis, band configurations, and surface transitions.
\newblock {\em Journal of Non-Newtonian Fluid Mechanics}, 151(1-3):101--118, 2008.

\bibitem{salipante2024two}
Paul~F Salipante, Michael Cromer, and Steven~D Hudson.
\newblock Two-species model for nonlinear flow of wormlike micelle solutions. part i: Model.
\newblock {\em Journal of Rheology}, 68(6):873--894, 2024.

\bibitem{varchanis2022evaluation}
Stylianos Varchanis, Simon~J Haward, Cameron~C Hopkins, John Tsamopoulos, and Amy~Q Shen.
\newblock Evaluation of constitutive models for shear-banding wormlike micellar solutions in simple and complex flows.
\newblock {\em Journal of Non-Newtonian Fluid Mechanics}, 307:104855, 2022.

\bibitem{lutz2016viscoelasticity}
Viviane Lutz-Bueno, Rossana Pasquino, Marianne Liebi, Joachim Kohlbrecher, and Peter Fischer.
\newblock Viscoelasticity enhancement of surfactant solutions depends on molecular conformation: Influence of surfactant headgroup structure and its counterion.
\newblock {\em Langmuir}, 32(17):4239--4250, 2016.

\bibitem{lutz2017intermicellar}
Viviane Lutz-Bueno, Marianne Liebi, Joachim Kohlbrecher, and Peter Fischer.
\newblock Intermicellar interactions and the viscoelasticity of surfactant solutions: Complementary use of sans and saxs.
\newblock {\em Langmuir}, 33(10):2617--2627, 2017.

\bibitem{zhang2018flow}
Yiran Zhang, Hadi Mohammadigoushki, Margaret~Y Hwang, and Susan~J Muller.
\newblock Flow of wormlike micellar fluids around a sharp bend: Effects of branching and shear-banding.
\newblock {\em Physical Review Fluids}, 3(9):093301, 2018.

\bibitem{gaudino2017effect}
Danila Gaudino, Rossana Pasquino, H~Kriegs, N~Szekely, W~Pyckhout-Hintzen, MP~Lettinga, and Nino Grizzuti.
\newblock Effect of the salt-induced micellar microstructure on the nonlinear shear flow behavior of ionic cetylpyridinium chloride surfactant solutions.
\newblock {\em Physical Review E}, 95(3):032603, 2017.

\bibitem{gaudino2015adding}
Danila Gaudino, Rossana Pasquino, and Nino Grizzuti.
\newblock Adding salt to a surfactant solution: Linear rheological response of the resulting morphologies.
\newblock {\em Journal of Rheology}, 59(6):1363--1375, 2015.

\bibitem{cheng2017distinguishing}
Peng Cheng, Michael~C Burroughs, L~Gary Leal, and Matthew~E Helgeson.
\newblock Distinguishing shear banding from shear thinning in flows with a shear stress gradient.
\newblock {\em Rheologica Acta}, 56:1007--1032, 2017.

\bibitem{al2020characterization}
Rehab~N Al-kaby, Sarah~L Codd, Joseph~D Seymour, and Jennifer~R Brown.
\newblock Characterization of velocity fluctuations and the transition from transient to steady state shear banding with and without pre-shear in a wormlike micelle solution under shear startup by rheo-nmr.
\newblock {\em Applied Rheology}, 30(1):1--13, 2020.

\bibitem{lopez2004shear}
MR~Lopez-Gonzalez, WM~Holmes, PT~Callaghan, and PJ~Photinos.
\newblock Shear banding fluctuations and nematic order in wormlike micelles.
\newblock {\em Physical review letters}, 93(26):268302, 2004.

\bibitem{lopez2006rheo}
MR~Lopez-Gonzalez, WM~Holmes, and PT~Callaghan.
\newblock Rheo-nmr phenomena of wormlike micelles.
\newblock {\em Soft Matter}, 2(10):855--869, 2006.

\bibitem{lerouge2008interface}
Sandra Lerouge, Marc-Antoine Fardin, M{\'e}d{\'e}ric Argentina, Guillaume Gr{\'e}goire, and Olivier Cardoso.
\newblock Interface dynamics in shear-banding flow of giant micelles.
\newblock {\em Soft Matter}, 4(9):1808--1819, 2008.

\bibitem{decruppe2006local}
Jean-Paul Decruppe, Olivier Greffier, S{\'e}bastien Manneville, and Sandra Lerouge.
\newblock Local velocity measurements in heterogeneous and time-dependent flows of a micellar solution.
\newblock {\em Physical Review E—Statistical, Nonlinear, and Soft Matter Physics}, 73(6):061509, 2006.

\bibitem{calabrese2015rheology}
Michelle~A Calabrese, Simon~A Rogers, Ryan~P Murphy, and Norman~J Wagner.
\newblock The rheology and microstructure of branched micelles under shear.
\newblock {\em Journal of Rheology}, 59(5):1299--1328, 2015.

\bibitem{calabrese2016understanding}
Michelle~A Calabrese, Simon~A Rogers, Lionel Porcar, and Norman~J Wagner.
\newblock Understanding steady and dynamic shear banding in a model wormlike micellar solution.
\newblock {\em Journal of Rheology}, 60(5):1001--1017, 2016.

\bibitem{thareja2011shear}
Prachi Thareja, Ingo~H Hoffmann, Matthew~W Liberatore, Matthew~E Helgeson, Y~Thomas Hu, Michael Gradzielski, and Norman~J Wagner.
\newblock Shear-induced phase separation (sips) with shear banding in solutions of cationic surfactant and salt.
\newblock {\em Journal of Rheology}, 55(6):1375--1397, 2011.

\bibitem{arenas2020alignment}
Brisa Arenas-G{\'o}mez, Cristina Garza, Yun Liu, and Rolando Castillo.
\newblock Alignment of worm-like micelles at intermediate and high shear rates.
\newblock {\em Journal of colloid and interface science}, 560:618--625, 2020.

\bibitem{cromer2014study}
Michael Cromer, Glenn~H Fredrickson, and L~Gary Leal.
\newblock A study of shear banding in polymer solutions.
\newblock {\em Physics of Fluids}, 26(6), 2014.

\bibitem{peterson2016shear}
Joseph~D Peterson, Michael Cromer, Glenn~H Fredrickson, and L~Gary~Leal.
\newblock Shear banding predictions for the two-fluid rolie-poly model.
\newblock {\em Journal of Rheology}, 60(5):927--951, 2016.

\bibitem{wu2021linear}
Shijian Wu and Hadi Mohammadigoushki.
\newblock Linear versus branched: flow of a wormlike micellar fluid past a falling sphere.
\newblock {\em Soft Matter}, 17(16):4395--4406, 2021.

\bibitem{tan2021determining}
Grace Tan, Weizhong Zou, Mike Weaver, and Ronald~G Larson.
\newblock Determining threadlike micelle lengths from rheometry.
\newblock {\em Journal of Rheology}, 65(1):59--71, 2021.

\bibitem{mohammadigoushki2018creeping}
Hadi Mohammadigoushki and Susan~J Muller.
\newblock Creeping flow of a wormlike micelle solution past a falling sphere: Role of boundary conditions.
\newblock {\em Journal of Non-Newtonian Fluid Mechanics}, 257:44--49, 2018.

\bibitem{yamamoto2009velocity}
Takehiro Yamamoto, Kazuhiro Sawa, and Kouki Mori.
\newblock Velocity measurements for shear flows of ctab/nasal aqueous solutions between parallel plates.
\newblock {\em Journal of Rheology}, 53(6):1347--1362, 2009.

\bibitem{becu2007evidence}
Lydiane Becu, Domitille Anache, Sebastien Manneville, and Annie Colin.
\newblock Evidence for three-dimensional unstable flows in shear-banding wormlike micelles.
\newblock {\em Physical Review E—Statistical, Nonlinear, and Soft Matter Physics}, 76(1):011503, 2007.

\bibitem{oelschlaeger2009linear}
C~Oelschlaeger, M~Schopferer, Frank Scheffold, and N~Willenbacher.
\newblock Linear-to-branched micelles transition: A rheometry and diffusing wave spectroscopy (dws) study.
\newblock {\em Langmuir}, 25(2):716--723, 2009.

\bibitem{galvan2008diffusing}
J~Galvan-Miyoshi, J~Delgado, and R~Castillo.
\newblock Diffusing wave spectroscopy in maxwellian fluids.
\newblock {\em The European Physical Journal E}, 26:369--377, 2008.

\bibitem{bales2002characterization}
Barney~L Bales and Raoul Zana.
\newblock Characterization of micelles of quaternary ammonium surfactants as reaction media i: dodeclytrimethylammonium bromide and chloride.
\newblock {\em The Journal of Physical Chemistry B}, 106(8):1926--1939, 2002.

\bibitem{buckingham1993effect}
Scott~A Buckingham, Christopher~J Garvey, and Gregory~G Warr.
\newblock Effect of head-group size on micellization and phase behavior in quaternary ammonium surfactant systems.
\newblock {\em The Journal of Physical Chemistry}, 97(39):10236--10244, 1993.

\bibitem{dhakal2016uniaxial}
Subas Dhakal and Radhakrishna Sureshkumar.
\newblock Uniaxial extension of surfactant micelles: counterion mediated chain stiffening and a mechanism of rupture by flow-induced energy redistribution.
\newblock {\em Acs Macro Letters}, 5(1):108--111, 2016.

\bibitem{schubert2003microstructure}
Beth~A Schubert, Eric~W Kaler, and Norman~J Wagner.
\newblock The microstructure and rheology of mixed cationic/anionic wormlike micelles.
\newblock {\em Langmuir}, 19(10):4079--4089, 2003.

\bibitem{lam2019structural}
Christopher~N Lam, Changwoo Do, Yangyang Wang, Guan-Rong Huang, and Wei-Ren Chen.
\newblock Structural properties of the evolution of ctab/nasal micelles investigated by sans and rheometry.
\newblock {\em Physical Chemistry Chemical Physics}, 21(33):18346--18351, 2019.

\bibitem{cates1987reptation}
M.~E. Cates.
\newblock Reptation of living polymers: Dynamics of entangled polymers in the presence of reversible chain-scission reactions.
\newblock {\em Macromolecules}, 20:2289--2296, 1987.

\bibitem{granek1992stress}
R~Granek and ME~Cates.
\newblock Stress relaxation in living polymers: Results from a poisson renewal model.
\newblock {\em The journal of chemical physics}, 96(6):4758--4767, 1992.

\bibitem{spenley1993nonlinear}
NA~Spenley, ME~Cates, and TCB McLeish.
\newblock Nonlinear rheology of wormlike micelles.
\newblock {\em Physical review letters}, 71(6):939, 1993.

\bibitem{burroughs2021flow}
Michael~C Burroughs, Yuanyi Zhang, Abhishek~M Shetty, Christopher~M Bates, L~Gary Leal, and Matthew~E Helgeson.
\newblock Flow-induced concentration nonuniformity and shear banding in entangled polymer solutions.
\newblock {\em Physical Review Letters}, 126(20):207801, 2021.

\bibitem{burroughs2023flow}
Michael~C Burroughs, Yuanyi Zhang, Abhishek Shetty, Christopher~M Bates, Matthew~E Helgeson, and L~Gary Leal.
\newblock Flow-concentration coupling determines features of nonhomogeneous flow and shear banding in entangled polymer solutions.
\newblock {\em Journal of Rheology}, 67(1):219--239, 2023.

\bibitem{castillo2011rheology}
J~Castillo-Tejas, JFJ Alvarado, S~Carro, F~P{\'e}rez-Villase{\~n}or, F~Bautista, and O~Manero.
\newblock Rheology of wormlike micelles from non-equilibrium molecular dynamics.
\newblock {\em Journal of Non-Newtonian Fluid Mechanics}, 166(3-4):194--207, 2011.

\bibitem{padding2008dynamics}
Johan~T Padding, Edo~S Boek, and Wim~J Briels.
\newblock Dynamics and rheology of wormlike micelles emerging from particulate computer simulations.
\newblock {\em The Journal of chemical physics}, 129(7), 2008.

\bibitem{boek2007flow}
ES~Boek, JT~Padding, VJ~Anderson, WJ~Briels, and JP~Crawshaw.
\newblock Flow of entangled wormlike micellar fluids: Mesoscopic simulations, rheology and $\mu$-piv experiments.
\newblock {\em Journal of non-newtonian fluid mechanics}, 146(1-3):11--21, 2007.

\end{thebibliography}
\end{document}